\pdfoutput=1
\documentclass[acmsmall,nonacm]{acmart}

\usepackage{multirow} 
\usepackage{multicol} 
\usepackage{booktabs}
\usepackage{amsthm}
\usepackage{algorithm}
\usepackage{algorithmic}

\usepackage{amsfonts}
\usepackage{bm}
\usepackage{amsmath}
\usepackage{graphicx} %
\usepackage{caption} %
\usepackage{mathrsfs}
\usepackage{subfigure}

\usepackage[table]{xcolor}
\usepackage{enumitem}
\usepackage{xspace}
\usepackage{comment}
\usepackage[noabbrev,capitalise]{cleveref}
\usepackage{balance}
\usepackage{mathtools}

\usepackage{ragged2e}

\usepackage{mdframed}
\usepackage{tcolorbox}

\usepackage{colortbl}
\usepackage{url}

\definecolor{lightblue}{RGB}{173,216,230}
\definecolor{lightgreen}{RGB}{144,238,144}

\newcommand{\ie}{\emph{i.e., }}
\newcommand{\eg}{\emph{e.g., }}

\newcommand{\nosection}[1]{\vspace{2pt}\noindent\textbf{#1.}}

\newcommand{\ourmodel}{LRPRec\xspace}

\AtBeginDocument{%
  }

\begin{document}

\title{Learning Robust Personalized Prompts for LLM-Driven Sequential Recommendation}

\author{Xiaolin Zheng}
\affiliation{%
  \institution{Zhejiang University}
  \city{Hangzhou}
  \country{China}
}
\email{xlzheng@zju.edu.cn}

\author{Qiyong Zhong}
\affiliation{%
  \institution{Zhejiang University}
  \city{Hangzhou}
  \country{China}
}
\email{youngzhong@zju.edu.cn}

\author{Jiajie Su}
\affiliation{%
  \institution{Zhejiang University}
  \city{Hangzhou}
  \country{China}
}
\email{sujiajie@zju.edu.cn}

\author{Xiang Chen}
\affiliation{%
  \institution{Zhejiang University}
  \city{Hangzhou}
  \country{China}
}
\email{wasdnsxchen@zju.edu.cn}

\renewcommand{\shortauthors}{Xiaolin Zheng et al.}

\begin{abstract}

With the rapid advancement of large language models (LLMs), LLM-driven sequential recommendation has emerged as a promising paradigm that formulates next-item prediction as autoregressive generation conditioned on natural-language prompts.
Despite their potential, existing methods exhibit high sensitivity to prompt formulations: even semantically equivalent prompts with minor wording variations can lead to substantial performance fluctuations, undermining robustness and incurring considerable manual engineering costs.
While continuous prompt learning offers a principled alternative by replacing discrete templates with learnable vectors, directly applying it to recommendation faces two interacting challenges:
(1) \emph{lack of personalization}, where existing continuous prompt methods learn a single task-level instruction shared across all users, yet recommendation inherently requires user-specific reasoning guidance;
and (2) \emph{semantic drift}, where continuous prompt vectors may be pushed by gradient updates into regions that the LLM cannot interpret meaningfully, degrading generation quality, a risk further amplified when user-specific signals are injected, as personalization pushes each user's prompt in a different direction away from the initialization.
To address these challenges, we propose \ourmodel, a learnable prompting framework tailored for LLM-driven sequential recommendation.
\ourmodel parameterizes instruction prompts as continuous vectors initialized from discrete templates and introduces two complementary mechanisms:
a \emph{personalized prompt injection} mechanism that encodes user behavioral signals into a preference embedding and additively injects it into the shared prompts, enabling user-level adaptation while maintaining parameter efficiency;
and a \emph{semantic drift constraint} that regularizes the shared instruction prompts to remain within a trust region around their initialization anchors, preventing optimization from departing the LLM's effective semantic space.
By applying the drift constraint to the shared component while personalization operates additively, our design decouples stability from expressiveness so that neither objective compromises the other.
Extensive experiments on three benchmark datasets demonstrate consistent improvements over strong baselines while eliminating the need for manual tuning of background and task inference templates.

\end{abstract}




\maketitle

\vspace{-0.08 in}
\section{Introduction}

Sequential recommendation (SR)~\cite{su2023personalized,hou2022towards,su2025distilling,speeder} aims to model users' evolving interests by leveraging their chronological interaction histories, and has become a fundamental component in modern recommender systems \cite{boka2024survey,wang2018attention,pctx}.
Early research in this area mainly adopts ID-based techniques \cite{rendle2010factorizing,tan2016improved}, which focus on capturing item transition patterns or co-occurrence signals from user behavior sequences.
Despite their effectiveness, such approaches are inherently limited when facing data sparsity and cold-start scenarios, where item identifiers alone provide insufficient semantic information \cite{cold-start-sess}.
To alleviate these limitations, content-aware SR methods have been proposed to incorporate auxiliary item information, including textual descriptions, visual content, and other multimodal features \cite{lai2022attribute,hou2022unisrec,hu2023adaptive}.
By enriching item representations with semantic cues beyond interaction logs, these methods attempt to enhance recommendation quality under sparse data conditions.
Nevertheless, their effectiveness is often restricted by the quality of multimodal representations, which are typically derived from task-agnostic or suboptimal pre-trained encoders and fail to fully capture fine-grained semantic relationships \cite{221,222}.

\begin{figure}[t]
\centering
\includegraphics[width=1.0\linewidth]{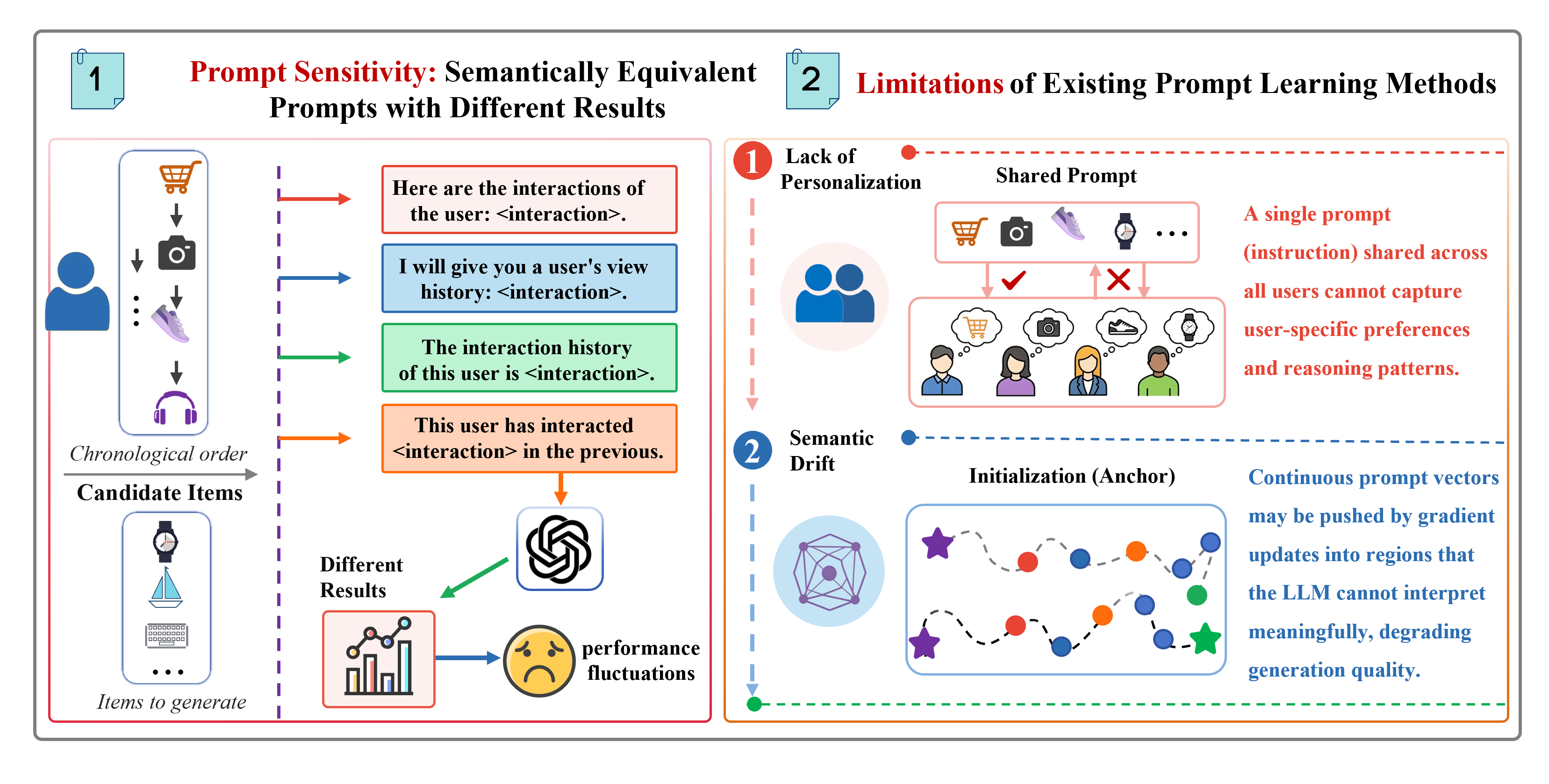}
\vspace{-0.05 in}
\caption{Figure 1. Motivation of \ourmodel. LLM-driven sequential recommendation suffers from prompt sensitivity, where minor wording variations in semantically equivalent prompts can lead to significant performance fluctuations. Although continuous prompt learning alleviates template dependence, existing methods still face two challenges: (1) lack of personalization, as a shared prompt cannot capture diverse user preferences and reasoning patterns, and (2) semantic drift, where learned prompt vectors may move away from the LLM’s effective semantic space during optimization.
}
\label{motivation}
\end{figure}

Recently, the rapid development of Large Language Models (LLMs) \cite{sod} has opened new opportunities for sequential recommendation, owing to their strong capabilities in semantic understanding, reasoning, and generation \cite{224,227}.
Existing LLM-based recommendation approaches can generally be divided into two paradigms \cite{222,224}: \emph{LLM-powered} methods, which utilize LLMs as content encoders to generate item representations consumed by conventional recommenders \cite{zhang2024notellm}, and \emph{LLM-driven} methods, which directly formulate recommendation as prompt-conditioned autoregressive generation \cite{ye2024harnessing,226,llara}.
While LLM-powered approaches benefit from expressive embeddings, they essentially treat LLMs as feature extractors without exploiting their generative capabilities, and often suffer from semantic misalignment between LLM-derived representations and downstream embedding spaces \cite{liu2024large,228}.
In contrast, LLM-driven methods leverage the intrinsic reasoning and language understanding of LLMs \cite{usa} to produce personalized results end-to-end, representing a more promising direction.
Representative methods include TALLRec~\cite{tallrec}, which reframes recommendation as instruction-tuned generation, and LLaRA~\cite{llara}, which combines textual prompts with collaborative embeddings projected from conventional recommenders.

However, as illustrated in Figure~\ref{motivation}, a key bottleneck of LLM-driven sequential recommendation lies in the heavy dependence on \emph{discrete natural-language prompts}.
The entire recommendation process is conditioned on a prompt that typically concatenates a \emph{background instruction}, a \emph{task description}, the \emph{interaction sequence}, and a \emph{candidate pool}.
We observe that existing methods are highly sensitive to the prompt formulation: even prompts that are semantically equivalent can produce substantially different rankings when small variations in wording, phrasing, or syntax are introduced (\eg reordering clauses, replacing synonyms, or changing punctuation).
Such prompt sensitivity undermines robustness and reproducibility, and forces practitioners to spend considerable effort on manual prompt engineering before deployment.
This issue is particularly problematic in SR, where the same system must reliably serve diverse users, one would expect consistent behavior as long as the essential information (interaction history and candidates) is provided, yet performance fluctuates noticeably across seemingly interchangeable templates.

Several prompt learning methods have been proposed to reduce the reliance on handcrafted discrete prompts.
POD~\cite{li2023pod} distills discrete prompts into continuous vectors to bridge user/item IDs and natural language.
LASER~\cite{zhang2024laser} inserts trainable virtual tokens at both the prefix and suffix of a frozen LLM to inject collaborative signals.
PPR~\cite{wu2024ppr} generates personalized prefix prompts conditioned on user profiles.
RPP~\cite{mao2025rpp} employs multi-agent reinforcement learning for instance-wise prompt assembly.
PromptAgent~\cite{wang2024promptagent} uses Monte Carlo Tree Search for automatic prompt optimization.
Among these, methods that operate in the discrete prompt space (\eg RPP, PromptAgent) still search over natural-language templates, meaning their outputs remain subject to the same wording sensitivity they aim to resolve.
In contrast, continuous prompt learning methods (\eg POD, LASER, PPR) eliminate template dependence by replacing discrete prompts with learnable vectors, representing a more principled direction.
However, directly applying continuous prompt tuning to recommendation still exhibits two notable limitations:
(1) \emph{Lack of user-level adaptive instruction prompts}: existing continuous prompt methods optimize task-level prompt representations shared across all users without providing user-specific adaptive instruction signals. While LASER injects collaborative signals via virtual tokens and PPR~\cite{wu2024ppr} introduces user-conditioned prefix generation, their personalization relies on shallow user profile features rather than leveraging the LLM's deep semantic understanding of behavioral patterns, limiting their ability to capture nuanced preference dynamics;
and (2) \emph{No control over semantic drift}: continuous prompt vectors can be pushed by gradients into regions that the LLM cannot interpret meaningfully, yet none of these methods provide explicit mechanisms to keep learned prompts within the LLM's effective semantic space.
As a result, existing continuous prompt learning approaches fall short of achieving both \emph{stable} and \emph{personalized} prompt optimization for LLM-driven SR.

To address these issues, we propose \textbf{\underline{L}}earning \textbf{\underline{R}}obust \textbf{\underline{P}}ersonalized Prompts for LLM-driven sequential \textbf{\underline{Rec}}ommendation (\ourmodel).
Our key insight is that while replacing discrete templates with learnable continuous prompts eliminates prompt sensitivity, naively applying standard prompt tuning to recommendation is insufficient due to two fundamental challenges unique to this setting.
Critically, these two challenges are not independent, they interact in a way that makes addressing them jointly non-trivial: injecting user-specific signals into prompts inherently increases the risk of semantic drift, as personalization pushes each user's prompt in a different direction away from the initialization; conversely, overly strict drift constraints can suppress the expressiveness needed for effective personalization.
Our framework is designed to navigate this tension.

\textbf{First, recommendation is inherently a personalization task.}
Standard prompt tuning learns a single task-level instruction shared across all instances, which is reasonable for generic NLP tasks where the objective is uniform (\eg ``translate this sentence'').
However, in sequential recommendation, different users exhibit vastly different preference patterns. For example, a user with concentrated interests requires different reasoning guidance than one with diverse tastes.
A fixed, instance-invariant prompt cannot provide optimal inference context for all users simultaneously.
To address this, we introduce a \emph{personalized prompt injection} mechanism that encodes user behavioral signals into a preference embedding and directly adds it to the shared instruction prompts, enabling user-level adaptation while maintaining parameter efficiency.

\textbf{Second, continuous prompts risk drifting away from the LLM's semantic space during optimization.}
Unlike discrete tokens that are inherently within the LLM's vocabulary \cite{masopd}, continuous vectors can be pushed by gradients into regions that the LLM cannot interpret meaningfully.
Our experiments confirm this: random initialization leads to severe degradation, demonstrating that prompts must reside in semantically valid regions.
While template-based initialization provides a good starting point, it offers no guarantee against drift during training, especially when personalized signals are injected, which amplifies the optimization dynamics.
To mitigate this, we propose a \emph{semantic drift constraint} that regularizes the shared instruction prompts to remain within a controlled neighborhood of their initialization anchors, ensuring that optimization improves upon the template without departing from the LLM's effective semantic space.

By applying the drift constraint specifically to the shared prompt component while allowing the personalized injection to operate additively, our design decouples the stability and personalization objectives, enabling both to be optimized without mutual interference.
This yields a robust and effective prompting framework that can be directly deployed by simply providing interaction histories and candidate pools.

Our contributions are summarized as follows:
\begin{itemize}[leftmargin=*]
\item We identify and empirically characterize the prompt sensitivity problem in LLM-driven sequential recommendation, showing that semantically equivalent prompts can cause substantial performance variance. We further analyze why standard continuous prompt tuning is insufficient for this setting, revealing two interacting challenges, lack of personalization and semantic drift that must be addressed jointly rather than in isolation.
\item We propose a personalized prompt injection mechanism that encodes user behavioral signals into a preference embedding and additively injects it into the shared learnable prompts, enabling user-level adaptation for the recommendation task while maintaining parameter efficiency.
\item We introduce a semantic drift constraint that regularizes the shared learnable prompts to remain within a trust region around their initialization anchors. By decoupling this constraint from the personalized component, our design prevents semantic drift without suppressing the expressiveness of user-specific adaptation.
\item Extensive experiments on three benchmark datasets demonstrate that \ourmodel consistently improves recommendation performance and robustness over strong baselines, while eliminating the need for manual tuning of background and task inference templates.
\end{itemize}

\section{Related Work}

\nosection{Sequential Recommendation}
Sequential recommendation has become a prominent research topic due to its effectiveness in modeling the dynamic evolution of user interests over time \cite{zheng2022ddghm,liu2023joint,su2023personalized}. Early studies typically relied on the Markov Chain assumption to capture sequential dependencies among user interactions \cite{rendle2010factorizing}. With the advancement of deep learning, a variety of neural architectures, including Recurrent Neural Networks (RNNs) \cite{wu2017recurrent}, Convolutional Neural Networks (CNNs) \cite{Caser}, and Graph Neural Networks (GNNs) \cite{chen2020handling}, have been introduced to model complex user--item interaction patterns. 
To alleviate inherent issues such as data sparsity and cold-start scenarios, content-aware sequential recommendation methods \cite{hou2022unisrec} incorporate auxiliary side information, including knowledge graphs \cite{huang2019explainable} and textual descriptions \cite{de2021transformers4rec,hou2022towards}, leading to substantial performance improvements. More recently, multimodal learning paradigms have further advanced representation learning in SR \cite{zhang2024multimodal,pan2022multimodal,ye2024harnessing}, enabling the capture of richer sequential behaviors across heterogeneous modalities \cite{ji2023online,hu2023adaptive}. Nevertheless, despite these improvements, conventional models remain limited in their ability to perform sophisticated semantic reasoning when compared with \textit{LLM-based} approaches \cite{222,225}.

\nosection{LLMs for Sequential Recommendation}
Owing to their strong representation learning and text generation capabilities \cite{llmsurvey}, Large Language Models (LLMs) have recently emerged as a promising direction for boosting sequential recommendation performance \cite{225,223}. Existing \textit{LLM-based} recommendation approaches can generally be categorized into two paradigms \cite{222,224,223}: \textit{LLM-powered} and \textit{LLM-driven} frameworks. 
The \textit{LLM-powered} paradigm \cite{zhang2024notellm,luo2024molar,tian2024mmrec} integrates LLMs into traditional recommender systems to enhance contextual understanding or semantic representations \cite{221}. For instance, NoteLLM \cite{zhang2024notellm} utilizes LLMs to generate compact note-level embeddings for recommendation refinement, while SAID \cite{hu2024enhancing} leverages LLMs to construct semantically consistent item representations. However, such methods fail to fully exploit the generative strengths of LLMs and often incur notable information loss when transforming LLM outputs into fixed embeddings for downstream tasks \cite{222,223}. 
In contrast, the \textit{LLM-driven} paradigm \cite{224,225,li2023e4srec,llara} directly formulates recommendation tasks as prompt-based generation problems, allowing LLMs to produce recommendation results end-to-end. Representative methods include TALLRec \cite{tallrec}, which reframes sequential recommendation as a binary prediction task conditioned on user histories, and LLaRA \cite{llara}, which combines textual prompts with collaborative embeddings projected from recommenders.

\nosection{Prompt Learning}
Recent studies have explored prompt learning~\cite{zhou2022learning,zhou2022conditional} as parameter-efficient paradigms to adapt pre-trained models for recommendation~\cite{yang2024graphpro,li2023personalized}, especially under data sparsity and cold-start settings. A foundational line of work reformulates recommendation as language processing with personalized prompt templates~\cite{geng2022recommendation,zhang2023instructrec}. Early continuous prompt tuning methods introduce learnable soft prompts to guide pre-trained encoders without full fine-tuning~\cite{ge2023domain,lester2021power}. POD~\cite{li2023pod} distills discrete prompts into continuous vectors to bridge user/item IDs and natural language, while LASER~\cite{zhang2024laser} proposes Bi-Tuning that inserts trainable virtual tokens at both the prefix and suffix of a frozen LLM to inject collaborative signals and enable controllable generation. Multi-stage approaches such as SCRec~\cite{dong2025screc} separately learn semantic and collaborative prompts before joint prediction~\cite{liu2023pre}. Beyond data-driven prompts, knowledge-enhanced methods such as KPT~\cite{zhai2023kpt} transform structured knowledge graphs into prompt representations to improve generalization~\cite{hu2022knowledgeable}. To address the inherent lack of collaborative filtering capability in LLMs, A-LLMRec~\cite{kim2024allMrec} injects pre-trained CF embeddings as prompt inputs, and Recommendation Critic~\cite{yang2025reccritic} trains a separate collaborative model whose outputs serve as soft prompts for the LLM. More recently, prompt learning has shifted toward user-level personalization: PPR~\cite{wu2024ppr} generates personalized prefix prompts from user profiles, PeaPOD~\cite{ramos2024peapod} distills individual preferences into soft prompts, CollaPPA~\cite{yue2025colappa} combines collaborative fine-tuning with personalized preference alignment prompts, and LLM-Rec~\cite{lyu2024llmrec} designs multiple prompting strategies for input text enrichment. Automatic prompt optimization methods, including RecPrompt~\cite{liu2024recprompt} with self-tuning prompt engineering for news recommendation, RPP~\cite{mao2025rpp} with multi-agent reinforcement learning for instance-wise prompt assembly, Chat-Rec~\cite{gao2023chatrec} with multi-turn dialogue prompts, and PromptAgent~\cite{wang2024promptagent} with Monte Carlo Tree Search, further enable the discovery of expert-level prompts beyond manual design~\cite{wood2023few}.

Unlike existing continuous prompt learning methods that learn task-level prompts without user-specific adaptive instruction signals and lack mechanisms to prevent semantic drift (\eg POD, LASER), \ourmodel uniquely combines personalized prompt injection with a semantic drift constraint. A key design consideration is that personalization and stability are inherently in tension: injecting user-specific signals increases drift risk, while strict drift control can suppress personalization. \ourmodel resolves this by applying the drift constraint to the shared prompt component while allowing personalization to operate additively, enabling both objectives to be achieved without mutual interference.

\section{Methodology}

\subsection{Problem Formulation}
\label{sec:prelim}

We study the sequential recommendation problem under the LLM-driven paradigm.
Let $\mathcal{I}$ denote the item universe.
For each user $u$, we observe a chronological interaction sequence
$\mathcal{S}_u = [i_{u,1}, i_{u,2}, \ldots, i_{u,T}]$, where $i_{u,t}\in\mathcal{I}$.
The goal is to predict the next interacted item $i_{u,T+1}$ based on $\mathcal{S}_u$.

In many LLM-driven recommender systems, the prediction is made over a candidate pool
$\mathcal{C}_u \subseteq \mathcal{I}$.
LLM-driven sequential recommendation reformulates next-item prediction as an autoregressive generation task.
Specifically, a prompt template (verbalization function) $\pi(\cdot)$ converts the structured inputs
$(\mathcal{S}_u, \mathcal{C}_u)$ into a token sequence (where we omit other prompts for simplicity):
\begin{equation}
\mathbf{X}_u \;=\; \pi(\mathcal{S}_u, \mathcal{C}_u).
\label{eq:prelim_prompt}
\end{equation}

Conditioned on $\mathbf{X}_u$, an LLM with parameters $\theta$ generates an output text sequence
$\mathbf{Y}_u$ in an autoregressive manner:
\begin{equation}
p_{\theta}(\mathbf{Y}_u \mid \mathbf{X}_u)
= \prod_{t=1}^{|\mathbf{Y}_u|}
p_{\theta}\!\left(Y_{u,t} \mid \mathbf{X}_u, \mathbf{Y}_{u,<t}\right).
\label{eq:prelim_ar}
\end{equation}

The generated output is then mapped to an item in $\mathcal{C}_u$ (\eg by matching an item title).

Despite its conceptual simplicity, this paradigm depends critically on the discrete natural-language prompt $\pi(\cdot)$.
In practice, even semantically equivalent prompts with minor variations in wording or syntax can lead to noticeable
performance fluctuations, which undermines robustness and incurs substantial manual prompt engineering costs.
This motivates us to explicitly factorize the prompt and to replace brittle discrete instruction templates
with a learnable prompting mechanism.

\subsection{An Overview of \ourmodel}
\label{sec:overview}

\ourmodel\ is a prompt-learning framework tailored for the LLM-driven sequential recommendation paradigm.
It targets a key practical limitation of prior LLM-driven methods: recommendation performance can fluctuate noticeably
under semantically equivalent natural-language prompts, which makes deployment brittle and incurs substantial manual
prompt engineering overhead.

Given a user interaction sequence $\mathcal{S}_u$ and a candidate pool $\mathcal{C}_u$, \ourmodel\ first constructs the
instance-specific \emph{evidence} via lightweight serialization (Eq.~\eqref{eq:evidence_ser}).
Instead of using discrete background and task templates, \ourmodel\ parameterizes the two \emph{instruction} components
as continuous learnable prompt vectors, \ie $\mathbf{P}^{\text{bg}}$ and $\mathbf{P}^{\text{task}}$ (Eq.~\eqref{eq:learnable_vectors}),
which are shared across all users and instances.
To adapt the shared prompts to individual users' preference patterns, we introduce a personalized prompt injection mechanism
that encodes user behavioral signals into a preference embedding and directly adds it to the instruction prompts (Section~\ref{sec:personalization}).
Furthermore, to prevent the learnable prompts from drifting away from the LLM's semantic space during optimization,
we incorporate a semantic drift constraint as a regularization term (Section~\ref{sec:drift_constraint}).

The LLM input is formed by concatenating the personalized instruction prompts with the evidence
(Eq.~\eqref{eq:final_input_personal}), and the LLM performs standard autoregressive generation to output the next-item prediction.
To make training efficient and practical, we first adapt the backbone LLM to the recommendation domain via parameter-efficient fine-tuning, and then freeze the adapted LLM weights entirely, updating only the learnable prompt vectors
and the projection parameters (Section~\ref{sec:training_objective}).
After training, the learned prompts and projection parameters are fixed and reused for all test instances, so inference only requires the raw
interaction sequence and candidate set as inputs (Section~\ref{sec:inference}), eliminating the need for manual tuning of background and task inference templates.

Overall, \ourmodel\ provides a principled replacement for discrete prompt templates in LLM-driven recommenders:
it learns a robust instruction context via continuous prompting, injects user-specific preference signals for personalization,
constrains shared prompt optimization within the LLM's semantic space, and enables stable recommendation under a parameter-efficient scheme.

\begin{figure*}[t]
\centering
\includegraphics[width=1\linewidth]{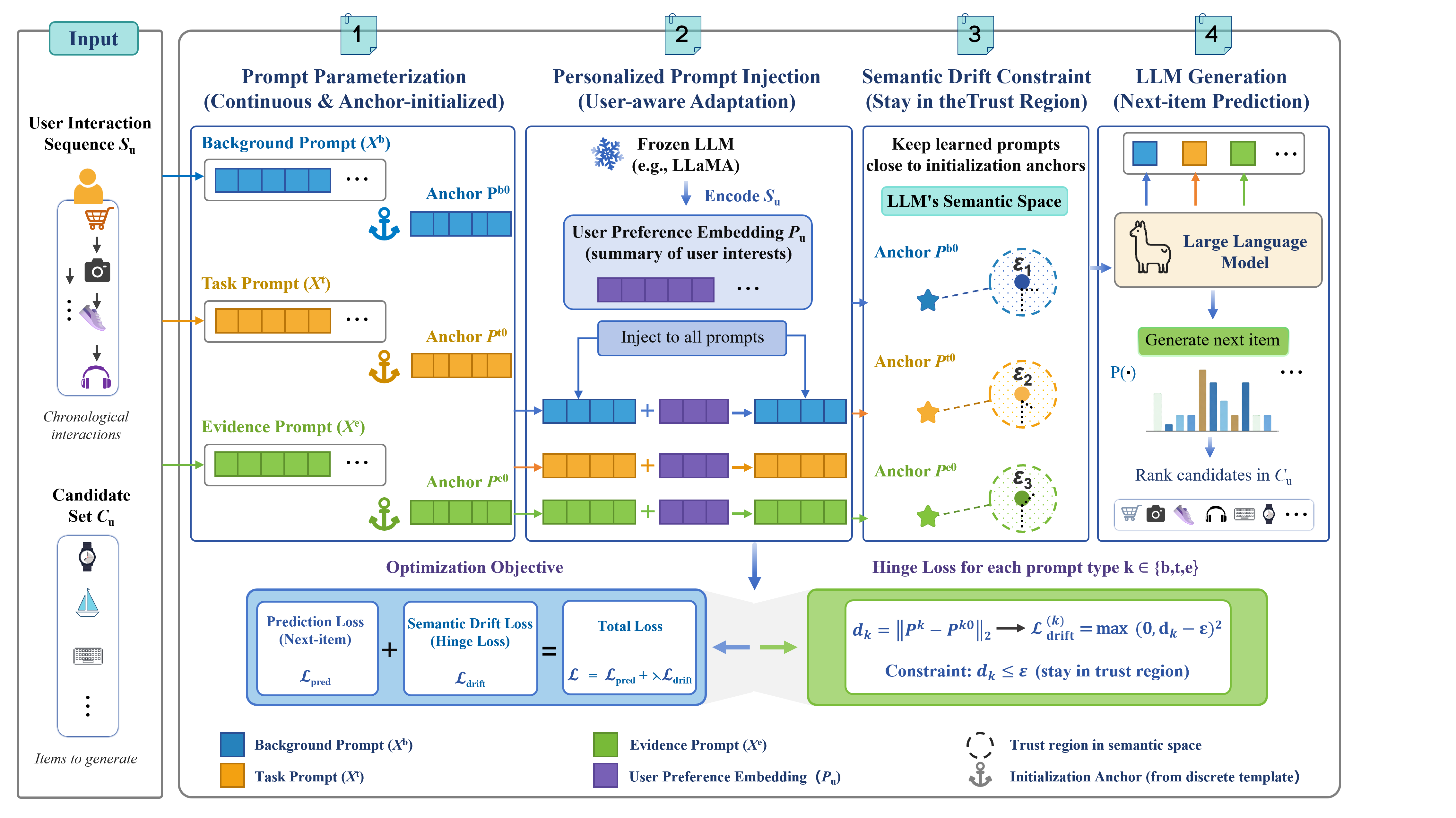}
\caption{The framework of \ourmodel. \ourmodel first parameterizes background, task, and evidence prompts as continuous vectors initialized from discrete template anchors. It then uses a frozen LLM to encode the user interaction sequence into a preference embedding, which is injected into the prompt vectors for user-aware adaptation. To prevent semantic drift, \ourmodel constrains learned prompts to remain within trust regions around their initialization anchors using a hinge-style drift loss. Finally, the personalized prompts guide the LLM to generate and rank the next-item prediction from the candidate set.}
\label{framework}
\end{figure*}

\subsection{Prompt Decomposition}
\label{sec:prompt_decomp}

A typical prompt used in LLM-driven recommendation interleaves two types of information:
(i) \emph{instruction}, which specifies what the model should do and how to format the answer,
and (ii) \emph{evidence}, which provides instance-specific inputs such as interaction histories and candidate sets.
To make this structure explicit, we decompose the prompt for user $u$ into four components:
\begin{equation}
\mathbf{X}_u
= \Big[\;
\mathbf{X}^{\text{bg}} \;;\;
\mathbf{X}^{\text{task}} \;;\;
\mathbf{X}^{\text{seq}}_u \;;\;
\mathbf{X}^{\text{cand}}_u
\;\Big],
\label{eq:prompt_decomp}
\end{equation}
where $\mathbf{X}^{\text{bg}}$ is a \emph{background prompt} providing global context,
$\mathbf{X}^{\text{task}}$ is a \emph{task prompt} describing the next-item prediction goal and the desired output format,
$\mathbf{X}^{\text{seq}}_u$ encodes the interaction sequence $\mathcal{S}_u$,
and $\mathbf{X}^{\text{cand}}_u$ encodes the candidate pool $\mathcal{C}_u$.
This decomposition aligns with the common prompt structure adopted in prior LLM-driven methods.

We treat $\mathbf{X}^{\text{bg}}$ and $\mathbf{X}^{\text{task}}$ as \emph{instance-invariant} components that are shared
across all examples, while $\mathbf{X}^{\text{seq}}_u$ and $\mathbf{X}^{\text{cand}}_u$ are \emph{instance-dependent}
evidence constructed from raw inputs.
In particular, we serialize the interaction sequence in chronological order and represent the candidate pool as:
\begin{equation}
\mathbf{X}^{\text{seq}}_u = \mathrm{Ser}(\mathcal{S}_u),\qquad
\mathbf{X}^{\text{cand}}_u = \mathrm{Ser}(\mathcal{C}_u),
\label{eq:evidence_ser}
\end{equation}
where $\mathrm{Ser}(\cdot)$ denotes a lightweight verbalization operator using item titles,
with delimiters to separate items.

This factorization isolates the prompt parts that primarily define \emph{instruction}
($\mathbf{X}^{\text{bg}}$ and $\mathbf{X}^{\text{task}}$) from those that provide \emph{evidence}
($\mathbf{X}^{\text{seq}}_u$ and $\mathbf{X}^{\text{cand}}_u$).
In the following, we keep the evidence construction unchanged, while learning robust instruction representations
to eliminate manual tuning of background and task templates and improve stability across prompt formulations.

\subsection{Learnable Prompting for Robust Recommendation}
\label{sec:learnable_prompt}

Based on the prompt decomposition in Eq.~\eqref{eq:prompt_decomp}, we observe that prompt sensitivity mainly originates from the \emph{instruction} components, \ie the discrete natural-language templates used for $\mathbf{X}^{\text{bg}}$ and $\mathbf{X}^{\text{task}}$.
In contrast, the evidence components $\mathbf{X}^{\text{seq}}_u$ and $\mathbf{X}^{\text{cand}}_u$ are directly derived from raw inputs and are naturally determined once the interaction history and candidate pool are given.
Therefore, we propose to keep the evidence construction unchanged and instead replace the brittle discrete instruction templates with \emph{learnable continuous prompts}.

We parameterize the background prompt and the task prompt as two sets of learnable vectors:
\begin{equation}
\mathbf{P}^{\text{bg}} = \{\mathbf{p}^{\text{bg}}_1, \ldots, \mathbf{p}^{\text{bg}}_{M}\}, \qquad
\mathbf{P}^{\text{task}} = \{\mathbf{p}^{\text{task}}_1, \ldots, \mathbf{p}^{\text{task}}_{N}\},
\label{eq:learnable_vectors}
\end{equation}
where $\mathbf{p}^{\text{bg}}_i, \mathbf{p}^{\text{task}}_j \in \mathbb{R}^{d}$ share the same dimensionality as the LLM input token embeddings, and $M$ and $N$ denote the prompt lengths determined by the number of tokens in the chosen initialization template.
Intuitively, $\mathbf{P}^{\text{bg}}$ provides a learnable global context for recommendation, while $\mathbf{P}^{\text{task}}$ specifies the learnable task instruction and output constraint.

\textbf{Initialization.}
We consider two practical initialization strategies.
(1) \emph{Template-based initialization}: initialize the learnable vectors from the token embeddings of a discrete prompt template (\eg a handcrafted background/task prompt).
(2) \emph{Random initialization}: randomly initialize all vectors with the same length as the chosen template.
In our experiments, the latter fails completely (discussed in Section~\ref{sec:ablation}), confirming that prompts must reside within the LLM's effective semantic space.
This observation motivates both our template-based initialization choice and the semantic drift constraint introduced in Section~\ref{sec:drift_constraint}.

While the above formulation eliminates the dependence on discrete templates, directly applying this instance-invariant prompting scheme to recommendation still faces two challenges, which we address in the following two subsections.

\subsection{Personalized Prompt Injection}
\label{sec:personalization}

The learnable prompts $\mathbf{P}^{\text{bg}}$ and $\mathbf{P}^{\text{task}}$ are shared across all users, providing a unified instruction context.
However, this instance-invariant design overlooks a fundamental characteristic of recommendation: \emph{personalization}.
In generic NLP tasks such as sentiment classification or machine translation, the task instruction is inherently uniform (\eg ``classify the sentiment of this review'' applies identically to all inputs).
In contrast, sequential recommendation requires the LLM to infer user-specific preferences from highly diverse behavioral patterns.
Users with concentrated interests (\eg exclusively watching horror movies) benefit from reasoning that emphasizes genre consistency, while users with broad tastes benefit from reasoning that captures recent interest transitions.
A single fixed instruction cannot optimally guide the LLM's inference for all these diverse scenarios.

To reconcile parameter efficiency with personalization, we propose \emph{personalized prompt injection}: the shared prompts serve as a repository of global recommendation knowledge, while a user-specific preference embedding is directly added to adapt them to each user's behavioral characteristics.

\textbf{User Preference Summarization.}
Rather than relying on shallow statistics (\eg mean-pooling token embeddings) that cannot capture high-level behavioral patterns, we leverage the frozen LLM itself to produce a semantically rich user preference representation.
Specifically, we construct a preference elicitation prompt that instructs the LLM to summarize the user's interests from the interaction history.
The prompt structure is illustrated below:

\begin{tcolorbox}[colframe=gray!30, colback=gray!5, coltitle=black!100, title=\textit{\textbf{User Preference Elicitation Prompt}}]
\textbf{\textless Instruction\textgreater:} Here is the user's interaction sequence: \textcolor{blue!70}{\textbf{[user\_history]}}.

Based on the above interaction history, summarize the user preference: \textcolor{red!70!black}{\textbf{\textless user\_preference\textgreater}}
\end{tcolorbox}

\noindent where \textbf{[user\_history]} is the serialized interaction sequence (\eg item titles in chronological order), and \textbf{\textless user\_preference\textgreater} is a designated special token appended at the end to serve as the preference aggregation anchor.

We feed the above prompt through the frozen LLM and extract the last hidden state at the \texttt{[user\_preference]} token position from the last transformer layer:
\begin{equation}
\mathbf{z}_u = \mathbf{h}^{(L)}_{\texttt{[user\_pref]}} \in \mathbb{R}^{d},
\label{eq:user_summary}
\end{equation}
where $L$ denotes the number of transformer layers.
Since the LLM processes the entire interaction sequence through its full depth before producing $\mathbf{h}^{(L)}_{\texttt{[user\_pref]}}$, this representation encodes a high-level semantic summary of the user's behavioral patterns, capturing not only item-level co-occurrence but also temporal dynamics and preference evolution that shallow aggregation methods cannot model.
Importantly, the LLM parameters remain completely frozen during this process; no gradients are propagated back through the LLM.
During training, we pre-compute $\mathbf{z}_u$ for all users offline and cache them, so this step introduces no additional computational overhead to the training loop.

\textbf{Preference Projection.}
Since the LLM's hidden-state representation and the learnable prompt vectors may not be semantically aligned, we project the user preference summary into the prompt space via a lightweight linear transformation:
\begin{equation}
\mathbf{e}_u = \mathbf{W}_p \mathbf{z}_u + \mathbf{b}_p \in \mathbb{R}^{d},
\label{eq:user_proj}
\end{equation}
where $\mathbf{W}_p \in \mathbb{R}^{d \times d}$ and $\mathbf{b}_p \in \mathbb{R}^{d}$ are learnable parameters.

\textbf{Direct Injection.}
The user preference embedding is directly added to each prompt vector, yielding personalized instruction prompts:
\begin{equation}
\tilde{\mathbf{p}}^{\text{bg}}_i = \mathbf{p}^{\text{bg}}_i + \mathbf{e}_u, \quad \forall i \in \{1, \ldots, M\},
\label{eq:inject_bg}
\end{equation}
\begin{equation}
\tilde{\mathbf{p}}^{\text{task}}_j = \mathbf{p}^{\text{task}}_j + \mathbf{e}_u, \quad \forall j \in \{1, \ldots, N\}.
\label{eq:inject_task}
\end{equation}

The final LLM input for user $u$ is then constructed as:
\begin{equation}
\widetilde{\mathbf{X}}_u
= \Big[\;
\tilde{\mathbf{P}}^{\text{bg}}_u \;;\;
\tilde{\mathbf{P}}^{\text{task}}_u \;;\;
\mathbf{X}^{\text{seq}}_u \;;\;
\mathbf{X}^{\text{cand}}_u
\;\Big],
\label{eq:final_input_personal}
\end{equation}
where $\tilde{\mathbf{P}}^{\text{bg}}_u = \{\tilde{\mathbf{p}}^{\text{bg}}_1, \ldots, \tilde{\mathbf{p}}^{\text{bg}}_M\}$ and $\tilde{\mathbf{P}}^{\text{task}}_u = \{\tilde{\mathbf{p}}^{\text{task}}_1, \ldots, \tilde{\mathbf{p}}^{\text{task}}_N\}$.

This design can be interpreted as decomposing each personalized prompt token $\tilde{\mathbf{p}}_i = \mathbf{p}_i + \mathbf{e}_u$ into a shared component $\mathbf{p}_i$ that encodes global recommendation knowledge learned from all training instances, and a user-specific component $\mathbf{e}_u$ that injects personalized preference signals.

\textbf{Rationale: Instruction-Level Personalization.}
A key insight behind this design is that, in the LLM's attention mechanism, instruction prompts directly modulate how the model attends to and reasons over the subsequent evidence tokens (\ie the interaction sequence and candidate pool).
By injecting user-specific signals into the instruction prompts themselves, rather than appending them as separate input segments, we effectively enable a single set of shared instruction templates to produce \emph{user-adaptive reasoning guidance}.
Concretely, for a user with concentrated interests, the injected preference embedding biases the instruction prompts toward emphasizing consistency patterns; for a user with diverse tastes, the same shared prompts are shifted toward capturing recent interest transitions.
This mechanism transforms the instruction prompts from static, universal directives into dynamic, user-conditioned reasoning controllers that steer the LLM's inference process differently for each individual, without increasing the input sequence length or requiring separate prompt parameters per user.

Compared with learning a separate prompt for each user (which would be prohibitively expensive) or appending user embeddings as additional input tokens (which increases sequence length and computational cost), our injection mechanism adds only a single $d \times d$ projection matrix and a bias vector, keeping the additional parameter count negligible relative to the LLM while achieving effective personalization at the instruction level.

\subsection{Semantic Drift Constraint}
\label{sec:drift_constraint}

In Section~\ref{sec:learnable_prompt}, we adopt template-based initialization to provide a semantically meaningful starting point for the learnable prompts.
The necessity of this choice is empirically confirmed by the failure of random initialization (Section~\ref{sec:ablation}), which demonstrates that prompts must reside within the LLM's effective semantic space to produce meaningful outputs.
However, template-based initialization only guarantees a valid starting point and provides no safeguard against the prompts gradually drifting into semantically invalid regions during gradient-based optimization.
As training progresses, the continuous prompt vectors may be pushed away from the manifold of meaningful token embeddings, causing the LLM to receive contexts it was never trained to interpret, ultimately degrading generation quality.

To address this, we introduce a \emph{semantic drift constraint} that regularizes the learned prompts to remain within a controlled neighborhood of their initialization anchors throughout training.

\textbf{Anchor Definition.}
We define the initialization embeddings as fixed semantic anchors:
\begin{equation}
\mathbf{A}^{\text{bg}} = \{\mathbf{a}^{\text{bg}}_1, \ldots, \mathbf{a}^{\text{bg}}_M\}, \quad
\mathbf{A}^{\text{task}} = \{\mathbf{a}^{\text{task}}_1, \ldots, \mathbf{a}^{\text{task}}_N\},
\label{eq:anchors}
\end{equation}
where each $\mathbf{a}^{\text{bg}}_i$ (resp.\ $\mathbf{a}^{\text{task}}_j$) is the token embedding used to initialize $\mathbf{p}^{\text{bg}}_i$ (resp.\ $\mathbf{p}^{\text{task}}_j$).
These anchors are determined at the beginning of training and remain fixed thereafter.

\textbf{Hinge-Style Regularization.}
We adopt a hinge-style constraint that allows prompts to freely optimize within a trust region of radius $\epsilon$, but penalizes deviations beyond this margin:
\begin{equation}
\mathcal{L}_{\text{drift}} = \frac{1}{M} \sum_{i=1}^{M} \max\!\left(0,\; \|\mathbf{p}^{\text{bg}}_i - \mathbf{a}^{\text{bg}}_i\|_2 - \epsilon\right)
+ \frac{1}{N} \sum_{j=1}^{N} \max\!\left(0,\; \|\mathbf{p}^{\text{task}}_j - \mathbf{a}^{\text{task}}_j\|_2 - \epsilon\right),
\label{eq:drift_loss}
\end{equation}
where $\epsilon > 0$ is a hyperparameter controlling the allowed drift radius.

The margin $\epsilon$ defines a trust region around each anchor: within this region, the prompt is free to improve upon the discrete template; beyond it, a linear penalty pulls the prompt back toward the anchor.
This design balances expressiveness (allowing the prompt to learn better representations than the original template) with stability (preventing drift into regions where the LLM cannot produce meaningful outputs).

\textbf{Discussion: Why Constrain Only Shared Prompts.}
One may note that the final input to the LLM is the personalized prompt $\tilde{\mathbf{p}}_i = \mathbf{p}_i + \mathbf{e}_u$ (Eqs.~\eqref{eq:inject_bg}--\eqref{eq:inject_task}), and question whether the drift constraint should also cover the user-specific component $\mathbf{e}_u$.
We argue that this is unnecessary for two reasons.
First, $\mathbf{e}_u$ is derived from the frozen LLM's own hidden-state representation (Eq.~\eqref{eq:user_summary}) via a lightweight linear projection (Eq.~\eqref{eq:user_proj}); since it originates from the LLM's internal semantic space, it is inherently interpretable to the model and does not face the same out-of-distribution risk as freely optimized continuous vectors.
Second, the semantic drift constraint targets a fundamentally different failure mode: it prevents the \emph{shared learnable parameters} $\mathbf{P}^{\text{bg}}, \mathbf{P}^{\text{task}}$ from being pushed by accumulated gradients into regions that no longer correspond to meaningful language concepts.
In contrast, $\mathbf{e}_u$ is not a learnable prompt vector but a \emph{data-dependent signal} computed fresh for each user instance; it adapts the shared prompts to individual preferences, which is precisely the intended personalization behavior.
Constraining $\mathbf{e}_u$ would suppress personalization and defeat the purpose of the injection mechanism.

\subsection{Training Objective and Optimization}
\label{sec:training_objective}

Let $\mathcal{D}=\{(\mathcal{S}_u,\mathcal{C}_u, i_{u,T+1})\}$ denote the training set.
For each instance, we define a target output text sequence $\mathbf{Y}^*_u$ that uniquely corresponds to the ground-truth next item $i_{u,T+1}$ (\eg its canonical title).
The recommendation loss optimizes the negative log-likelihood of generating $\mathbf{Y}^*_u$ conditioned on the personalized prompt input:
\begin{equation}
\mathcal{L}_{\text{rec}} = - \sum_{u} \log p_{\theta}\!\left(\mathbf{Y}^*_u \,\middle|\, \tilde{\mathbf{P}}^{\text{bg}}_u, \tilde{\mathbf{P}}^{\text{task}}_u, \mathbf{X}^{\text{seq}}_u, \mathbf{X}^{\text{cand}}_u \right).
\label{eq:rec_loss}
\end{equation}
Here, $p_{\theta}(\cdot)$ follows the autoregressive factorization in Eq.~\eqref{eq:prelim_ar}.
Since $\theta$ is frozen, gradients are applied only to the learnable prompt vectors $\mathbf{P}^{\text{bg}}$, $\mathbf{P}^{\text{task}}$ and the projection parameters $\mathbf{W}_p$, $\mathbf{b}_p$.

The overall training objective combines recommendation loss with the semantic drift constraint:
\begin{equation}
\mathcal{L} = \mathcal{L}_{\text{rec}} + \alpha \cdot \mathcal{L}_{\text{drift}},
\label{eq:total_loss}
\end{equation}
where $\alpha > 0$ is a balancing coefficient.
Note that $\mathcal{L}_{\text{drift}}$ only regularizes the shared prompt vectors $\mathbf{P}^{\text{bg}}$ and $\mathbf{P}^{\text{task}}$; the user-specific embedding $\mathbf{e}_u$ is instance-dependent and does not require drift regularization (see the Remark at the end of Section~\ref{sec:drift_constraint} for a detailed justification).

All learnable parameters ($\mathbf{P}^{\text{bg}}$, $\mathbf{P}^{\text{task}}$, $\mathbf{W}_p$, $\mathbf{b}_p$) are jointly optimized under this objective.
The semantic drift constraint ensures that the prompts remain within the LLM's effective semantic space throughout training,
providing a spatial safeguard that prevents optimization from departing meaningful regions.

\textbf{Decoupled Two-Stage Training.}
A prerequisite for effective prompt optimization is that the backbone LLM already possesses basic recommendation capabilities.
If the learnable prompts are trained on top of a general-purpose LLM that has never been exposed to recommendation data, the prompts must simultaneously teach the model \emph{what recommendation is} and \emph{how to optimally perform it}, two fundamentally different objectives that interfere with each other during optimization.
To decouple these concerns, we adopt a two-stage training procedure:
\begin{itemize}[leftmargin=*]
\item \textbf{Stage~1: Recommendation Adaptation.}
We first fine-tune the backbone LLM on the recommendation task using LoRA~\cite{hu2022lora}, a parameter-efficient adaptation method.
During this stage, we use standard discrete prompt templates and train the model with the same next-item prediction objective (Eq.~\eqref{eq:rec_loss}), allowing the LLM to acquire domain-specific knowledge such as item co-occurrence patterns and sequential transition dynamics.
\item \textbf{Stage~2: Prompt Optimization.}
After Stage~1 converges, we load the adapted checkpoint, freeze \emph{all} LLM parameters (including the LoRA weights), and train only the learnable prompt vectors $\mathbf{P}^{\text{bg}}$, $\mathbf{P}^{\text{task}}$ and the projection parameters $\mathbf{W}_p$, $\mathbf{b}_p$ under the full objective $\mathcal{L}$ (Eq.~\eqref{eq:total_loss}).
\end{itemize}
This decoupled design offers two advantages.
First, it allows the prompt optimization to focus exclusively on learning optimal instruction representations, since the backbone already understands the recommendation task.
Second, it avoids gradient interference between the LLM adaptation and prompt optimization objectives, leading to more stable convergence.

\subsection{Inference}
\label{sec:inference}

After training, the learned prompts $\mathbf{P}^{\text{bg}}$, $\mathbf{P}^{\text{task}}$ and the projection parameters $\mathbf{W}_p$, $\mathbf{b}_p$ are fixed and reused across all test instances.
Given a new user sequence $\mathcal{S}_u$ and candidate pool $\mathcal{C}_u$, we construct the model input as follows:
(1) compute the user preference representation $\mathbf{z}_u$ by performing a single forward pass of the frozen LLM on the preference elicitation prompt (Eq.~\eqref{eq:user_summary}), then obtain the preference embedding $\mathbf{e}_u$ via Eq.~\eqref{eq:user_proj};
(2) inject it into the shared prompts via Eqs.~\eqref{eq:inject_bg}--\eqref{eq:inject_task};
(3) concatenate with the serialized evidence:
\begin{equation}
\widetilde{\mathbf{X}}_u
= \Big[\;
\tilde{\mathbf{P}}^{\text{bg}}_u \;;
\tilde{\mathbf{P}}^{\text{task}}_u \;;
\mathrm{Ser}(\mathcal{S}_u) \;;
\mathrm{Ser}(\mathcal{C}_u)
\;\Big],
\label{eq:test_input}
\end{equation}
and perform LLM decoding to obtain the predicted output $\widehat{\mathbf{Y}}_u$.

The final recommendation $\hat{i}_{u,T+1}$ is obtained by mapping $\widehat{\mathbf{Y}}_u$ back to an item in the candidate pool.
Notably, inference requires no manual tuning of background or task instruction templates: the system only needs the raw interaction sequence and candidate set as inputs,
while the instruction context is provided by the learned continuous prompts personalized with user behavioral signals.

\textbf{Efficiency Considerations.}
The preference representation $\mathbf{z}_u$ can be cached for each user and only needs to be recomputed when the user's interaction sequence is updated.
For users whose sequences remain unchanged since training, the cached $\mathbf{z}_u$ from the offline pre-computation stage can be directly reused.
For new users or users with updated sequences, a single LLM forward pass (without autoregressive decoding) is required to obtain $\mathbf{z}_u$.
Since this forward pass processes only the preference elicitation prompt (not the full recommendation input), its cost is modest compared to the subsequent autoregressive decoding step, making deployment practical.

\section{Experiments and Analysis}

\subsection{Experimental Setup}

\nosection{Datasets} We evaluate on three datasets:
\begin{itemize}[leftmargin=*]
    \item \textbf{MovieLens} \cite{movielens} is a widely adopted benchmark for movie recommendation, which provides explicit user ratings together with corresponding movie titles.
    \item \textbf{Steam} \cite{SASRec} contains user-generated reviews of video games from the Steam Store, along with the associated game titles.
    \item \textbf{LastFM} \cite{lastfm}, sourced from the Last.fm music service, records user--artist listening interactions as well as artist names.
\end{itemize}
Since fine-tuning large language models typically incurs substantially higher computational costs compared to training conventional recommendation models, we preprocess and sample datasets to keep the scale tractable. Specifically, we retain the MovieLens dataset at its original scale, and for the Steam dataset, we uniformly sample one-third of the users and one-third of the games, preserving the interactions among them to construct a dataset of moderate size. Across all datasets, interaction sequences are ordered by timestamp and split into training, validation, and test sets with a ratio of 8:1:1. This temporal splitting strategy ensures that future interactions are excluded from the training phase, effectively preventing information leakage \cite{data_leakage}. After preprocessing, MovieLens contains 1,478 items, 1,063 user sequences, and 112,384 interactions; Steam contains 3,987 items, 13,426 user sequences, and 241,653 interactions; LastFM contains 5,142 items, 1,387 user sequences, and 83,917 interactions.

\nosection{Baseline Algorithms} We compare \ourmodel with the following representative baselines spanning four categories.

\begin{itemize}[leftmargin=*]
    \item \textbf{Traditional Sequential Recommenders.}
    We include several representative sequential recommendation models that differ in their underlying architectures.
    \textbf{GRU4Rec} \cite{GRU4Rec} models user behavior sequences using gated recurrent units, capturing sequential dependencies through recurrent state transitions.
    \textbf{Caser} \cite{Caser} leverages convolutional neural networks to extract local and global patterns from interaction sequences by applying horizontal and vertical convolutional filters.
    \textbf{SASRec} \cite{SASRec} adopts a self-attention mechanism to model long-range dependencies among items, and has become a strong and widely used attention-based baseline for sequential recommendation.
    \textbf{HGN}~\cite{hgn} introduces a hierarchical gating network that captures both user-level and item-level sequential patterns through feature gating and instance gating mechanisms.
    \textbf{BERT4Rec}~\cite{sun2019bert4rec} employs a bidirectional transformer with masked item prediction to model users' interaction preferences, capturing left-to-right and right-to-left sequential dependencies.

    \item \textbf{Proprietary LLM APIs.}
    We evaluate several state-of-the-art proprietary large language models accessed via API in a zero-shot setting.
(1) \textbf{GPT-5.4}~\cite{singh2025openai} is a contemporary proprietary LLM from OpenAI, demonstrating strong reasoning and instruction-following capabilities across diverse tasks.
    (2) \textbf{Claude-Opus-4.6} is Anthropic's most capable model, excelling at complex reasoning, nuanced instruction following, and long-context understanding.
    (3) \textbf{Gemini-3.1-Flash}~\cite{team2023gemini} is Google's efficient multimodal model optimized for low-latency inference while maintaining strong generation quality.
    (4) \textbf{DeepSeek-V4-Flash}~\cite{deepseek2026deepseek} is a highly efficient open-weight model from DeepSeek-AI with million-token context support and competitive reasoning performance.

    \item \textbf{LLM-driven Sequential Recommenders.}
    We further compare our approach with LLM-based recommenders that are fine-tuned or adapted for recommendation tasks.
    (1) \textbf{Llama2-7B}~\cite{llama2} is a widely used open-source large language model released by Meta, which has demonstrated strong capabilities in instruction following and text generation.
    (2) \textbf{Llama3.1-8B-Instruct}~\cite{grattafiori2024llama} is an instruction-tuned variant of the Llama 3.1 family, offering improved instruction following and reasoning capabilities over its predecessor.
    (3) \textbf{MoRec}~\cite{MoRec_or_IDRec} augments conventional recommendation models by incorporating item-side modality information, such as textual features, through a dedicated text encoder. 
    Following the official implementation, we instantiate MoRec with BERT as the text encoder and SASRec as the recommendation backbone.
    (4) \textbf{TALLRec}~\cite{tallrec} applies instruction tuning to large language models using recommendation-oriented corpora, enabling LLMs to perform recommendation tasks via natural language prompts.
    Since TALLRec formulates recommendation as a binary decision problem (\ie predicting \emph{YES} or \emph{NO} for a target item), we adapt it to our setting by reformulating the task as selecting the next item from a given candidate set.
    (5) \textbf{LLaRA}~\cite{llara} is a Large Language-Recommendation Assistant framework for sequential recommendation that integrates a conventional sequential recommender with a large language model to perform next-item prediction.
    Following the official SASRec-based implementation, we instantiate LLaRA with SASRec as the sequential recommendation backbone, and represent each item using a hybrid prompt that concatenates item text tokens with a behavioral token projected from the SASRec item embedding via an adapter.
    (6) \textbf{MLLM-MSR}~\cite{ye2024harnessing} harnesses multimodal large language models for multimodal sequential recommendation by encoding visual and textual item features jointly.

    \item \textbf{Prompt Learning Enhanced Methods.}
    We compare with prompt-based approaches that optimize soft or discrete prompts for LLM-based recommendation.
(1) \textbf{POD}~\cite{li2023pod} distills the recommendation knowledge of large LLMs into compact continuous prompt representations for efficient inference.
(2) \textbf{PPR}~\cite{wu2024ppr} generates personalized soft prefix prompts conditioned on user profiles to adapt a frozen LLM for sequential recommendation.
    (3) \textbf{LASER}~\cite{zhang2024laser} introduces a parameter-efficient bi-tuning strategy that inserts learnable virtual tokens at both the input and output ends of an LLM to inject collaborative signals.
    (4) \textbf{PromptAgent}~\cite{wang2024promptagent} formulates prompt optimization as a strategic planning problem and employs Monte Carlo tree search to automatically discover expert-level prompts.
(5) \textbf{RPP}~\cite{mao2025rpp} leverages reinforcement learning to personalize discrete prompt components on a per-user basis, enabling instance-wise prompt selection for recommendation.

\end{itemize}

\nosection{Evaluation Metrics}
Following LLaRA~\cite{llara} and MLLM-MSR~\cite{ye2024harnessing}, we adopt a candidate-based evaluation protocol with a candidate set size of 20 (1 ground-truth item + 19 uniformly sampled negatives that the user has not previously interacted with).
Because (M)LLM-based methods embed item descriptions (\eg titles) directly into the prompt, an excessively large candidate pool would exceed the context window or cause out-of-memory issues during training and inference.
For fair comparison, \ourmodel and all baselines that incorporate a candidate pool in the prompt use the same pool size of 20.
Each method is required to identify the correct next item from this pool, and we assess performance using the HitRatio@1 (HR@1) metric, which is computed only over valid predictions (\ie outputs that correspond to items in the candidate pool): $\text{HR@1} = \text{\# correct predictions} / \text{\# valid outputs}$.
When properly prompted, LLM-based recommendation models are able to directly output a single predicted item.
In contrast, for conventional recommendation approaches, we treat the item with the highest predicted score as the final recommendation.
Notably, as \ourmodel adopts a generative prediction framework, it may occasionally produce invalid outputs, such as meaningless tokens or items that do not belong to the predefined candidate pool.
To account for this issue, we further report an auxiliary metric, termed ValidRatio, which measures the fraction of predictions that correspond to valid candidate items across all sequences, thereby reflecting the models' ability to follow task instructions.
Inspired by~\cite{speeder}, some models may achieve a high \textit{HitRatio@1}, yet fail to generate valid or meaningful recommendations. 
To mitigate this issue, we introduce a new evaluation metric, Valid HitRatio@1 (VHR@1), defined as $VHR@1 = ValidRatio \times HitRatio@1$.
This composite metric penalizes models that achieve high accuracy only among valid outputs while producing a large fraction of invalid predictions, thereby preventing inflated performance estimates from models that ``gamble'' on a subset of valid generations.
By jointly accounting for both validity and accuracy in a single score, VHR@1 provides a more faithful reflection of real-world deployment quality than reporting the two metrics independently.

\nosection{Implementation Details}
\label{sec:impl_details}
(1) Backbone LLM.
We use Llama3.1-8B-Instruct~\cite{grattafiori2024llama} as the backbone.
The dimension of the hidden states is 4096.
(2) Prompt construction.
Following Eq.~\eqref{eq:prompt_decomp}, each input consists of four components:
(i) the background prompts, (ii) the task prompts, (iii) the interaction sequence, and (iv) the candidate pool.
The interaction sequence $\mathbf{X}^{\text{seq}}_u$ is serialized in chronological order, and each item is represented by
its title.
The candidate pool $\mathbf{X}^{\text{cand}}_u$ is serialized as a list with explicit delimiters.
(3) Learnable prompts.
We parameterize the instruction prompts as continuous learnable vectors
$\mathbf{P}^{\text{bg}}$ and $\mathbf{P}^{\text{task}}$ (Eq.~\eqref{eq:learnable_vectors}).
We adopt template-based initialization, where each learnable vector is initialized from the corresponding token embedding of a
handcrafted prompt template.
The prompt lengths $M$ and $N$ are naturally determined by the number of tokens in the chosen initialization template; in our default setting, the background template tokenizes to $M=12$ tokens and the task template tokenizes to $N=26$ tokens.
(4) Personalized prompt injection.
For the user preference projection (Eq.~\eqref{eq:user_proj}), we employ a single linear layer with input and output dimension $d=4096$ to project the user preference summary into the prompt space.
(5) Semantic drift constraint.
For the drift constraint (Eq.~\eqref{eq:drift_loss}), the trust-region radius is set to $\epsilon=0.5$ and the balancing coefficient to $\alpha=0.1$ by default.
(6) Training procedure.
As described in Section~\ref{sec:training_objective}, \ourmodel\ adopts a decoupled two-stage training procedure.
In Stage~1 (recommendation adaptation), we fine-tune the backbone LLM using LoRA~\cite{hu2022lora} with rank $r=8$ and scaling factor $\alpha_{\text{lora}}=16$, applied to the query and value projection matrices.
We use standard discrete prompt templates and train for 5 epochs with a learning rate of $2 \times 10^{-4}$, a batch size of 128, and a cosine learning rate scheduler with warm-up.
In Stage~2 (prompt optimization), we load the Stage~1 checkpoint, freeze all LLM parameters (including LoRA weights), and train only the learnable prompt vectors and projection parameters.
We employ a warm-up strategy,
initiated with 1/100 of the maximum learning rate, and adjust it using a cosine scheduler.
The maximum learning rate is set to $5 \times 10^{-4}$.
The batch size is set to $B=128$, and gradient clipping is applied with a maximum norm of $1.0$.
All LLM-driven sequential recommenders and prompt learning enhanced methods are trained for 10 epochs, and we select the checkpoint that achieves the best performance on the validation set for final evaluation on the test set.
Our implementations for conventional recommenders follow~\cite{llara}, employing the Adam optimizer with a learning rate of 0.001, an embedding dimension of 64, and a batch size of 256.
All experiments are conducted on 8 NVIDIA H20 GPUs (96\,GB memory each).
To mitigate the impact of randomness, all experiments are conducted 5 times with distinct random seeds, and the results are averaged.

\begin{table*}[t]
\renewcommand\arraystretch{0.95}
\centering
\caption{Experimental results on three datasets. The best results are boldfaced and the second-best results are underlined. ``$*$'' denotes statistically significant improvements ($p < 0.05$) over the best baseline according to a paired t-test.}
\label{tab:compare}
\resizebox{0.95\textwidth}{!}{
\begin{tabular}{c|ccc|ccc|ccc}
\toprule
\multirow{2}{*}{\textbf{Methods}}
& \multicolumn{3}{c|}{\textbf{MovieLens}} 
& \multicolumn{3}{c|}{\textbf{Steam}} 
& \multicolumn{3}{c}{\textbf{LastFM}} \\
\cmidrule(lr){2-4} \cmidrule(lr){5-7} \cmidrule(lr){8-10}
& HR@1 & ValidRatio & VHR@1 & HR@1 & ValidRatio & VHR@1 & HR@1 & ValidRatio & VHR@1 \\
\midrule
\rowcolor[HTML]{FEE090}
\multicolumn{10}{c}{\textbf{ID-based Sequential Recommendation}} \\[4pt]
\textbf{Caser} & 0.3924 & 1.0000 & 0.3924 & 0.4287 & 1.0000 & 0.4287 & 0.2312 & 1.0000 & 0.2312 \\
\textbf{GRU4Rec} & 0.3807 & 1.0000 & 0.3807 & 0.4093 & 1.0000 & 0.4093 & 0.2538 & 1.0000 & 0.2538 \\
\textbf{HGN} & 0.3987 & 1.0000 & 0.3987 & 0.4412 & 1.0000 & 0.4412 & 0.2689 & 1.0000 & 0.2689 \\
\textbf{SASRec} & 0.3523 & 1.0000 & 0.3523 & 0.3936 & 1.0000 & 0.3936 & 0.2156 & 1.0000 & 0.2156 \\
\textbf{BERT4Rec} & 0.3926 & 1.0000 & 0.3926 & 0.4495 & 1.0000 & 0.4495 & 0.2721 & 1.0000 & 0.2721 \\
\midrule
\rowcolor[HTML]{E0F3F8}
\multicolumn{10}{c}{\textbf{Proprietary LLM APIs}} \\[4pt]
\textbf{GPT-5.4} & 0.3842 & 0.6237 & 0.2396 & 0.4183 & 0.5842 & 0.2444 & 0.4126 & 0.5963 & 0.2460 \\
\textbf{Claude-Opus-4.6} & 0.3796 & 0.6014 & 0.2283 & 0.4127 & 0.5716 & 0.2359 & 0.4078 & 0.5834 & 0.2379 \\
\textbf{Gemini-3.1-Flash} & 0.3518 & 0.5683 & 0.1999 & 0.3876 & 0.5247 & 0.2034 & 0.3814 & 0.5412 & 0.2064 \\
\textbf{DeepSeek-V4-Flash} & 0.3247 & 0.5321 & 0.1728 & 0.3612 & 0.4918 & 0.1776 & 0.3563 & 0.5087 & 0.1812 \\
\midrule
\rowcolor[HTML]{D5E8D4}
\multicolumn{10}{c}{\textbf{LLM-driven Sequential Recommendation}} \\[4pt]
\textbf{Llama2-7B} & 0.0437 & 0.4518 & 0.0197 & 0.0128 & 0.1587 & 0.0020 & 0.0238 & 0.3521 & 0.0084 \\
\textbf{Llama3.1-8B-Instruct} & 0.1847 & 0.6312 & 0.1166 & 0.2156 & 0.5473 & 0.1180 & 0.1423 & 0.5847 & 0.0832 \\
\textbf{MoRec} & 0.2716 & 1.0000 & 0.2716 & 0.3827 & 1.0000 & 0.3827 & 0.1713 & 1.0000 & 0.1713 \\
\textbf{TALLRec} & 0.3978 & 0.9318 & 0.3707 & 0.4572 & 0.9873 & 0.4514 & 0.4263 & 0.9791 & 0.4174 \\
\textbf{LLaRA} & 0.4356 & 0.9721 & 0.4234 & 0.5012 & 0.9948 & 0.4986 & 0.4437 & 1.0000 & 0.4437 \\
\textbf{MLLM-MSR} & \underline{0.4523} & 0.9823 & \underline{0.4443} & 0.5063 & 0.9891 & 0.5008 & \underline{0.4627} & 0.9867 & \underline{0.4565} \\
\midrule
\rowcolor[HTML]{FFF2CC}
\multicolumn{10}{c}{\textbf{Prompt Learning Enhanced Methods}} \\[4pt]
\textbf{POD} & 0.4253 & 0.9742 & 0.4143 & 0.4812 & 0.9918 & 0.4773 & 0.4361 & 0.9912 & 0.4323 \\
\textbf{PPR} & 0.4187 & 0.9658 & 0.4044 & 0.4756 & 0.9863 & 0.4691 & 0.4289 & 0.9845 & 0.4223 \\
\textbf{LASER} & 0.4487 & 0.9812 & 0.4403 & 0.5036 & 0.9953 & 0.5012 & 0.4589 & 0.9934 & 0.4559 \\
\textbf{PromptAgent} & 0.4312 & 0.9587 & 0.4134 & 0.4868 & 0.9826 & 0.4783 & 0.4417 & 0.9778 & 0.4319 \\
\textbf{RPP} & 0.4412 & 0.9763 & 0.4307 & \underline{0.5078} & 0.9941 & \underline{0.5048} & 0.4568 & 0.9901 & 0.4523 \\
\midrule
\rowcolor[HTML]{F8CECC}
\textbf{\ourmodel} & \textbf{0.4947}$^*$ & \textbf{1.0000} & \textbf{0.4947}$^*$ & \textbf{0.5412}$^*$ & \textbf{1.0000} & \textbf{0.5412}$^*$ & \textbf{0.5016}$^*$ & \textbf{1.0000} & \textbf{0.5016}$^*$ \\
\bottomrule
\end{tabular}
}
\end{table*}

\subsection{Overall Performance}

\label{overall_performance}

We evaluate \ourmodel against baselines from four categories on MovieLens, Steam, and LastFM. The results are reported in Table~\ref{tab:compare} in terms of HR@1, ValidRatio, and VHR@1. From the results, we derive the following observations:

\textbf{(1) Well-adapted LLM methods outperform ID-based methods, but adaptation quality matters.}
Well-adapted LLM-driven recommenders (\eg TALLRec, LLaRA, MLLM-MSR) and prompt learning enhanced methods generally outperform ID-based sequential recommenders across all datasets, as shown in Table~\ref{tab:compare}. However, vanilla or weakly adapted LLM variants (\eg Llama2-7B, Llama3.1-8B-Instruct, MoRec) may underperform ID-based baselines, indicating that the reasoning power of LLMs can only be effectively harnessed through proper task-specific adaptation.

\textbf{(2) Fine-tuning is essential for LLMs on recommendation tasks.}
Models without task-specific fine-tuning exhibit notably lower VHR@1 compared to fine-tuned counterparts. In particular, directly prompting LLMs without adaptation leads to extremely low ValidRatio, while even powerful proprietary models cannot match fine-tuned methods. This confirms that task-specific adaptation is critical for effective recommendation, as general-purpose LLMs often generate invalid or out-of-domain items without proper alignment.
It is worth noting that proprietary LLM APIs achieve moderate HR@1 among their valid outputs (\eg GPT-5.4 attains HR@1 $\approx$ 0.41 on LastFM), which appears competitive with fine-tuned methods.
However, their low ValidRatio ($\approx$ 0.60) means that nearly 40\% of all predictions are unusable, the model either generates items outside the candidate pool or produces malformed outputs.
This makes these models unreliable for real-world deployment where every user query must receive a valid recommendation.
Moreover, proprietary API calls incur substantial monetary costs and latency overhead that scale linearly with the number of users and requests, rendering them impractical for production-scale recommender systems that must serve millions of users in real time.

\textbf{(3) \ourmodel outperforms all baselines.}
Among all methods, \ourmodel consistently achieves the best performance across all three datasets. Notably, \ourmodel maintains a perfect ValidRatio of 1.0000 on all datasets, indicating that the learned prompts reliably guide the LLM to produce valid candidate items. Compared with the strongest baseline, \ourmodel achieves notable relative improvements in VHR@1 across all datasets. These improvements stem from \ourmodel's personalized learnable prompting mechanism, which injects user-specific preference signals into the instruction context while constraining semantic drift, thereby providing more effective and stable guidance for the LLM's generation process.

\begin{table*}[t]
\renewcommand\arraystretch{0.9}
\centering
\caption{Ablation study on key components of \ourmodel. ``w/o'' denotes removing a component; ``w/'' denotes replacing with an alternative.}
{\fontsize{6.8}{7.5}\selectfont
\label{ablation}
\resizebox{0.98\linewidth}{!}{
\begin{tabular}{c|@{\hskip 4pt}c@{\hskip 4pt}c@{\hskip 4pt}c@{\hskip 4pt}|@{\hskip 4pt}c@{\hskip 4pt}c@{\hskip 4pt}c@{\hskip 4pt}|@{\hskip 4pt}c@{\hskip 4pt}c@{\hskip 4pt}c}
\toprule
\multirow{2}{*}{\textbf{Variants}}
& \multicolumn{3}{c|}{\textbf{MovieLens}}
& \multicolumn{3}{c|}{\textbf{Steam}}
& \multicolumn{3}{c}{\textbf{LastFM}} \\
\cmidrule(lr){2-4} \cmidrule(lr){5-7} \cmidrule(lr){8-10}
& HR@1 & ValidRatio & VHR@1
& HR@1 & ValidRatio & VHR@1
& HR@1 & ValidRatio & VHR@1 \\
\midrule

\rowcolor[HTML]{FEE090}
\multicolumn{10}{c}{\textbf{Core Module Ablation}} \\[2pt]

\textbf{w/o Personalization}
& 0.4512 & 0.9983 & 0.4504
& 0.5087 & 0.9967 & 0.5070
& 0.4612 & 1.0000 & 0.4612 \\

\textbf{w/o Drift Constraint}
& 0.4723 & 0.9647 & 0.4556
& 0.5198 & 0.9712 & 0.5048
& 0.4836 & 0.9623 & 0.4654 \\

\midrule
\rowcolor[HTML]{E0F3F8}
\multicolumn{10}{c}{\textbf{Initialization \& Prompt Strategy}} \\[2pt]

\textbf{w/ Random Initialization}
& 0.0875 & 0.5198 & 0.0455
& 0.0922 & 0.5673 & 0.0523
& 0.0769 & 0.5328 & 0.0410 \\

\textbf{w/ Discrete Prompt}
& 0.4589 & 0.9712 & 0.4457
& 0.5067 & 0.9935 & 0.5034
& 0.4587 & 1.0000 & 0.4587 \\

\midrule
\rowcolor[HTML]{D5E8D4}
\multicolumn{10}{c}{\textbf{Personalization Design Alternatives}} \\[2pt]

\textbf{w/ Mean-Pool Preference}
& 0.4687 & 1.0000 & 0.4687
& 0.5213 & 0.9978 & 0.5202
& 0.4756 & 1.0000 & 0.4756 \\

\textbf{w/ Concat Injection}
& 0.4823 & 1.0000 & 0.4823
& 0.5287 & 0.9989 & 0.5281
& 0.4878 & 1.0000 & 0.4878 \\

\midrule

\textbf{\ourmodel}
& \textbf{0.4947} & \textbf{1.0000} & \textbf{0.4947}
& \textbf{0.5412} & \textbf{1.0000} & \textbf{0.5412}
& \textbf{0.5016} & \textbf{1.0000} & \textbf{0.5016} \\

\bottomrule
\end{tabular}
}}
\vspace{0.15 in}
\end{table*}

\subsection{Ablation Study}
\label{sec:ablation}

To validate the contribution of each component in \ourmodel, we design six ablation variants grouped into three categories: (1) core module ablation, (2) initialization and prompt strategy, and (3) personalization design alternatives.
Results are reported in Table~\ref{ablation}.

\subsubsection{\textbf{w/o Personalization}}
This variant removes the personalized prompt injection mechanism entirely.
The shared prompts $\mathbf{P}^{\text{bg}}$ and $\mathbf{P}^{\text{task}}$ are used directly without adding the user preference embedding $\mathbf{e}_u$, meaning all users receive identical instruction prompts.
As shown in Table~\ref{ablation}, removing personalization leads to consistent and notable performance drops across all three datasets.
This confirms that a single fixed instruction context cannot optimally guide the LLM's inference for users with diverse behavioral patterns.
The personalized injection mechanism enables the shared prompts to produce user-adaptive reasoning guidance, which is essential for recommendation quality.
Notably, the ValidRatio remains high in this variant, indicating that the learnable prompts still maintain valid generation behavior. The performance loss is purely in recommendation accuracy, underscoring that personalization contributes to \emph{precision} rather than \emph{validity}.

\subsubsection{\textbf{w/o Drift Constraint}}
This variant removes the semantic drift constraint by setting $\alpha = 0$ in Eq.~\eqref{eq:total_loss}, allowing the learnable prompts to optimize freely without any regularization toward their initialization anchors.
As shown in Table~\ref{ablation}, we observe two effects: (1) ValidRatio drops noticeably, indicating that unconstrained optimization pushes prompts into regions where the LLM produces more invalid outputs; (2) despite maintaining reasonable HitRatio@1, the overall VHR@1 degrades due to the compounding effect of lower validity.
These results validate our hypothesis that gradient-based optimization can gradually drift the prompt vectors away from the LLM's effective semantic space.
The hinge-style constraint provides a necessary spatial safeguard: it allows the prompts to improve upon the discrete template within a trust region while preventing them from entering semantically invalid regions that the LLM was never trained to interpret.

\subsubsection{\textbf{w/ Random Initialization}}
This variant replaces the template-based initialization with random initialization, \ie both $\mathbf{P}^{\text{bg}}$ and $\mathbf{P}^{\text{task}}$ are initialized from a random distribution.
As shown in Table~\ref{ablation}, this leads to severe degradation: VHR@1 drops to below 6\% on all datasets, while ValidRatio also substantially decreases.
This demonstrates that the LLM's generation behavior is extremely sensitive to the semantic validity of its input context.
Random vectors lie far outside the manifold of meaningful token embeddings, causing the LLM to produce incoherent outputs regardless of subsequent optimization.
This result provides strong empirical evidence for our design choice of template-based initialization and further motivates the semantic drift constraint: if the starting point matters this much, preventing drift away from it during training is equally critical.

\subsubsection{\textbf{w/ Discrete Prompt}}
\begin{figure*}[t]
\centering
\includegraphics[width=0.98\linewidth]{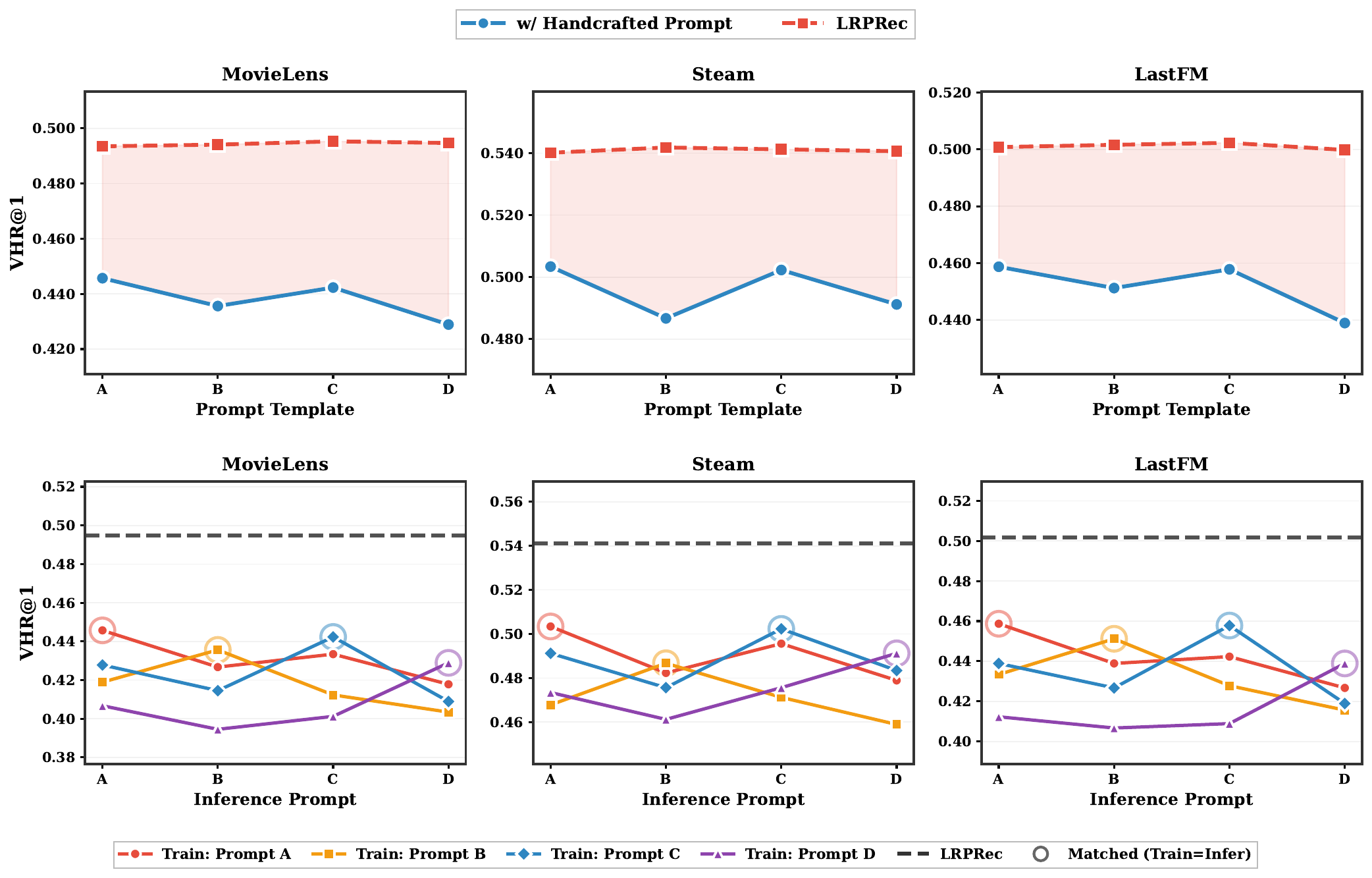}
\caption{Generalization analysis. Top row: Prompt sensitivity, showing VHR@1 of handcrafted prompts (blue solid) vs.\ \ourmodel\ (red dashed) across four templates on each dataset; the shaded area highlights the performance gap. Bottom row: Cross-template transfer, where each colored line represents a model trained with one template and evaluated under all four templates at inference time; the black dashed line denotes \ourmodel, which is template-agnostic; circled markers indicate matched (train=infer) conditions.}
\label{vis}
\end{figure*}

This variant discards learnable prompting entirely and instead uses the best-performing handcrafted discrete prompt template identified through manual prompt engineering on each dataset.
As shown in Table~\ref{ablation}, while discrete prompts achieve reasonable performance, they consistently underperform \ourmodel\ across all datasets.
Moreover, discrete prompts exhibit high variance across different template formulations (as visualized in Figure~\ref{vis}): semantically equivalent templates with minor wording differences can lead to notable performance fluctuations.
In contrast, \ourmodel\ learns a single robust instruction context that automatically adapts to the backbone LLM and data distribution, eliminating the need for manual trial-and-error and providing stable performance.

\subsubsection{\textbf{w/ Mean-Pool Preference}}
This variant replaces our LLM-based user preference summarization with a simpler mean-pooling approach: instead of feeding the interaction sequence through the frozen LLM and extracting the hidden state at the \texttt{[user\_preference]} token, we directly compute the mean of the token embeddings from the interaction sequence as the user representation $\mathbf{z}_u$.
As shown in Table~\ref{ablation}, this leads to moderate but consistent performance drops across all datasets.
The mean-pooling approach can only capture shallow token co-occurrence statistics, while our LLM-based summarization leverages the full depth of the transformer to encode high-level behavioral patterns, including temporal dynamics and preference evolution.
This validates the benefit of using the LLM's deep representations for user preference modeling, even though the LLM parameters remain frozen.

\subsubsection{\textbf{w/ Concat Injection}}
This variant changes the injection mechanism: instead of directly adding $\mathbf{e}_u$ to each prompt token, we prepend $\mathbf{e}_u$ as an additional prefix token before the instruction prompts, \ie $\widetilde{\mathbf{X}}_u = [\mathbf{e}_u; \mathbf{P}^{\text{bg}}; \mathbf{P}^{\text{task}}; \mathbf{X}^{\text{seq}}_u; \mathbf{X}^{\text{cand}}_u]$.
As shown in Table~\ref{ablation}, this approach achieves competitive but slightly lower performance than \ourmodel\ across all datasets.
The concatenation approach relies on the LLM's attention mechanism to implicitly route user preference information to the instruction tokens, which is less direct than our additive injection.
Moreover, concatenation increases the input sequence length, adding computational overhead during both training and inference.
Our additive injection provides a more efficient and effective instruction-level personalization by directly modulating the prompt vectors in the embedding space.

\subsection{Generalization Analysis}
\label{sec:generalization}

To evaluate whether \ourmodel\ can alleviate the brittleness caused by discrete prompt formulations, we conduct a prompt-level generalization study.
Specifically, we prepare four handcrafted prompt templates, denoted as \texttt{Prompt\_A}--\texttt{Prompt\_D}, which are semantically equivalent but differ in wording, syntactic structure, and length.
All templates contain the same essential components (a background role description, the interaction sequence, the candidate pool, and an explicit selection instruction) but present them with different phrasing and organization.
We design these templates to vary along three dimensions: \emph{verbosity} (concise vs.\ detailed), \emph{formality} (conversational vs.\ technical), and \emph{structural organization} (flat narrative vs.\ explicitly labeled sections), thereby covering a representative range of prompt styles that practitioners might employ in practice.
The four prompts used in this study are as follows:

\textbf{Prompt\_A:}
\texttt{You are a recommendation system. Here are the interactions of the user: <Interaction>. Here is the candidate pool: <Candidate>. Please choose exactly one item from the candidate pool as the next item the user will interact with. Only output the chosen item.}

\textbf{Prompt\_B:}
\texttt{I will give you a user's interaction history and a set of candidate items. User's interaction history: <Interaction>. Candidate items: <Candidate>. Based on the user's history, select the most likely next item from the candidate items. Return only the item name.}

\textbf{Prompt\_C:}
\texttt{The user has previously interacted with the following items: <Interaction>. Now you are given a list of candidate items: <Candidate>. Task: predict which candidate item the user will interact with next. Answer with one item from the candidate list.}

\textbf{Prompt\_D:}
\texttt{You are an expert sequential recommender. Input: [Interaction Sequence] <Interaction> [Candidate Pool] <Candidate>. Instruction: Pick the single best next-item prediction from the candidate pool. Your output must be exactly one candidate item, without explanation or additional text.}

For each dataset, we use each of the four templates as the initialization source for \ourmodel's learnable prompts.
Concretely, the token embeddings of a given template are extracted from the LLM's embedding layer and used to initialize the learnable vectors $\mathbf{P}^{\text{bg}}$ and $\mathbf{P}^{\text{task}}$ (Eq.~\eqref{eq:learnable_vectors}).
Since the four templates differ in length, the resulting prompt lengths $M$ and $N$ naturally vary across initialization templates.
This yields four independently trained \ourmodel\ instances per dataset, each starting from a different semantic anchor.
The goal is to examine whether the final learned prompts converge to similar performance regardless of the initialization template, thereby validating that \ourmodel\ is robust to the choice of handcrafted template used for initialization.

\textbf{Initialization Robustness.}
We first evaluate each trained model under two inference protocols, as shown in the top row of Figure~\ref{vis}.
\textbf{(1) w/ handcrafted prompt:} we directly use the same discrete template (\eg \texttt{Prompt\_A}) at inference time for the model trained with that template.
The corresponding results are shown as solid lines in the top row of Figure~\ref{vis}.
\textbf{(2) \ourmodel\ (full model at inference):} for each trained model, we apply the complete \ourmodel\ inference pipeline following Eq.~\eqref{eq:test_input}: we compute the user preference embedding $\mathbf{e}_u$ from the frozen LLM, inject it into the learned continuous instruction vectors $\mathbf{P}^{\text{bg}}$ and $\mathbf{P}^{\text{task}}$ to obtain personalized prompts $\tilde{\mathbf{P}}^{\text{bg}}_u$ and $\tilde{\mathbf{P}}^{\text{task}}_u$, and concatenate them with the instance evidence (interaction sequence and candidate pool).
Importantly, this inference does not rely on any handcrafted template beyond the evidence serialization.
The results are shown as dashed lines in the top row of Figure~\ref{vis}.

Across MovieLens, Steam, and LastFM, we highlight two consistent patterns from the top row of Figure~\ref{vis}:
\textbf{(1) Discrete prompts exhibit high sensitivity to template wording.} The solid lines fluctuate noticeably across \texttt{Prompt\_A}--\texttt{Prompt\_D}, confirming that the performance of LLM-driven recommendation is sensitive to the specific discrete prompt used during training and inference.
\textbf{(2) \ourmodel\ achieves robust and superior performance regardless of initialization template.} The dashed lines are substantially more stable with respect to the initialization template choice and remain consistently above the solid lines for all four templates on all datasets, demonstrating both improved robustness and stronger performance.
These results indicate that regardless of which handcrafted template is used to initialize the learnable prompts, the optimization process guided by the semantic drift constraint converges to a high-quality prompt representation within the LLM's effective semantic space.
The personalized prompt injection further contributes to this robustness: by conditioning the instruction prompts on user behavioral signals, \ourmodel\ anchors its reasoning in user-specific preference patterns rather than relying on the surface-level wording of discrete templates, enabling stable and high-quality recommendation without manual tuning of inference-time instruction templates.

\textbf{Cross-Template Transfer.}
Beyond evaluating models with their matched training templates, we further investigate a more challenging scenario: \emph{what happens when a model trained with one discrete template is evaluated using a different template at inference time?}
This cross-template transfer experiment reveals the degree to which discrete-prompt-based models overfit to specific template wordings.
As shown in the bottom row of Figure~\ref{vis}, for each dataset we train four models (one per template) and evaluate each model under all four templates at inference time, yielding a $4 \times 4$ transfer matrix.

We summarize three key findings from the bottom row of Figure~\ref{vis}:
\textbf{(1) Matched templates yield the best discrete-prompt performance.} All discrete-prompt models achieve their best performance under the matched condition (train template = inference template), indicated by the circled markers. This confirms that these models partially memorize surface-level template patterns rather than learning purely task-relevant representations.
\textbf{(2) Mismatched templates cause consistent degradation.} The performance degradation under mismatched conditions is consistent across all datasets. Notably, the severity of transfer degradation varies across models: some templates produce more transferable representations while others are more sensitive to template mismatch, indicating that this brittleness is pervasive and unpredictable in practice.
\textbf{(3) \ourmodel\ surpasses all transfer conditions.} \ourmodel\ (black dashed line) consistently outperforms \emph{all} transfer conditions, including the matched diagonal, across every dataset. This demonstrates that \ourmodel\ does not merely match the best-case discrete prompt performance but fundamentally transcends the limitations of template-dependent inference.

These results have important practical implications: in real-world deployment, the inference prompt may need to be modified (\eg for A/B testing, multilingual adaptation, or system updates), and discrete-prompt models would suffer unpredictable performance degradation.
\ourmodel\ eliminates this fragility entirely by replacing template-dependent inference with learned, template-agnostic continuous instructions.

\subsection{Hyperparameter Sensitivity Analysis}
\label{sec:sensitivity}

We investigate the sensitivity of \ourmodel\ to two key hyperparameters: the drift constraint radius $\epsilon$ and the constraint weight $\alpha$.
For each experiment, we vary one hyperparameter while keeping the other at its default value ($\epsilon=0.5$, $\alpha=0.1$).
Results are reported at VHR@1 across all datasets.

\subsubsection{\textbf{Effect of Drift Radius $\epsilon$}}

The drift radius $\epsilon$ controls the size of the trust region around each initialization anchor.
We vary $\epsilon \in \{0.1, 0.3, 0.5, 0.7, 1.0, 2.0\}$ and report results in Figure~\ref{fig:sensitivity}(a).

When $\epsilon$ is too small (\eg 0.1), the prompts are overly constrained and cannot deviate sufficiently from the discrete template to learn improved representations, resulting in performance close to the discrete prompt baseline (w/ Discrete Prompt in Table~\ref{ablation}).
As $\epsilon$ increases to the range of 0.3--0.7, performance improves substantially and remains stable, indicating that the prompts have sufficient freedom to optimize while still being anchored within the LLM's semantic space.
Notably, the per-dataset optimum varies slightly within this range (\eg Steam peaks at $\epsilon=0.3$ while LastFM peaks at $\epsilon=0.7$), but the differences are marginal.
Beyond $\epsilon = 1.0$, performance begins to degrade, as the constraint becomes too loose to effectively prevent semantic drift.
At $\epsilon = 2.0$, the behavior approaches the unconstrained variant (w/o Drift Constraint), confirming that an excessively large trust region provides diminishing regularization.
We select $\epsilon = 0.5$ as the default value because it achieves consistently strong performance across all datasets without requiring per-dataset tuning.

\subsubsection{\textbf{Effect of Constraint Weight $\alpha$}}

\begin{figure*}[t]
\centering
\includegraphics[width=0.95\linewidth]{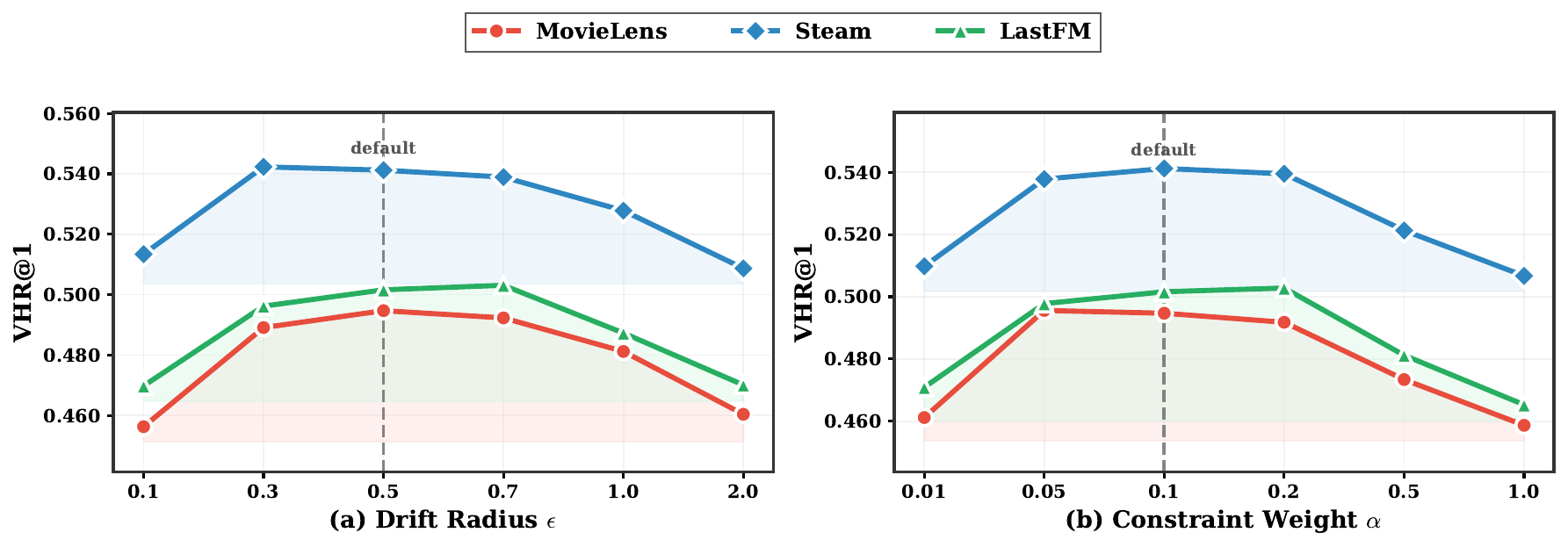}
\caption{Hyperparameter sensitivity analysis of \ourmodel\ on three datasets. (a) Drift radius $\epsilon$. (b) Constraint weight $\alpha$. The dashed vertical lines indicate the default values used in all other experiments.}
\label{fig:sensitivity}
\end{figure*}

The constraint weight $\alpha$ balances the recommendation loss $\mathcal{L}_{\text{rec}}$ and the drift penalty $\mathcal{L}_{\text{drift}}$.
We vary $\alpha \in \{0.01, 0.05, 0.1, 0.2, 0.5, 1.0\}$ in Figure~\ref{fig:sensitivity}(b).

When $\alpha$ is very small (\eg 0.01), the drift constraint has negligible effect, and the model behaves similarly to the unconstrained variant.
As $\alpha$ increases to 0.05--0.2, performance improves and stabilizes, indicating that moderate regularization strength is sufficient to keep prompts within the effective semantic space.
Similar to $\epsilon$, the per-dataset optimum within this plateau varies slightly (\eg MovieLens favors $\alpha=0.05$ while LastFM favors $\alpha=0.2$), but the differences remain negligible.
However, when $\alpha$ becomes too large (\eg 0.5 or 1.0), the constraint dominates the optimization objective, severely suppressing the model's ability to learn task-specific prompt representations and leading to significant performance degradation on all datasets.
This sharp degradation at high $\alpha$ values contrasts with the drift radius $\epsilon$, which exhibits a more gradual performance change: while the model is robust within $\alpha \in [0.05, 0.2]$, practitioners should avoid over-regularization.
We adopt $\alpha = 0.1$ as the default since it provides the best trade-off across all three datasets.

\subsection{Cold-Start Analysis}
\label{sec:coldstart}

A key advantage of \ourmodel\ is its personalized prompt injection mechanism, which encodes user behavioral signals into a preference embedding $\mathbf{e}_u$ and injects it into the shared learnable prompts.
A natural question arises: \emph{how does this mechanism perform when user interaction history is sparse?}
To answer this, we conduct a cold-start analysis by partitioning test users into two groups based on their available interaction history length:

\begin{itemize}[leftmargin=*]
\item \textbf{Cold} ($\leq$3 interactions): Users with very sparse behavioral records. The preference embedding is extracted from this short interaction sequence.
\item \textbf{Warm} ($>$3 interactions): Users with sufficient history, representing the standard setting.
\end{itemize}

\begin{figure}[t]
\centering
\includegraphics[width=0.98\linewidth]{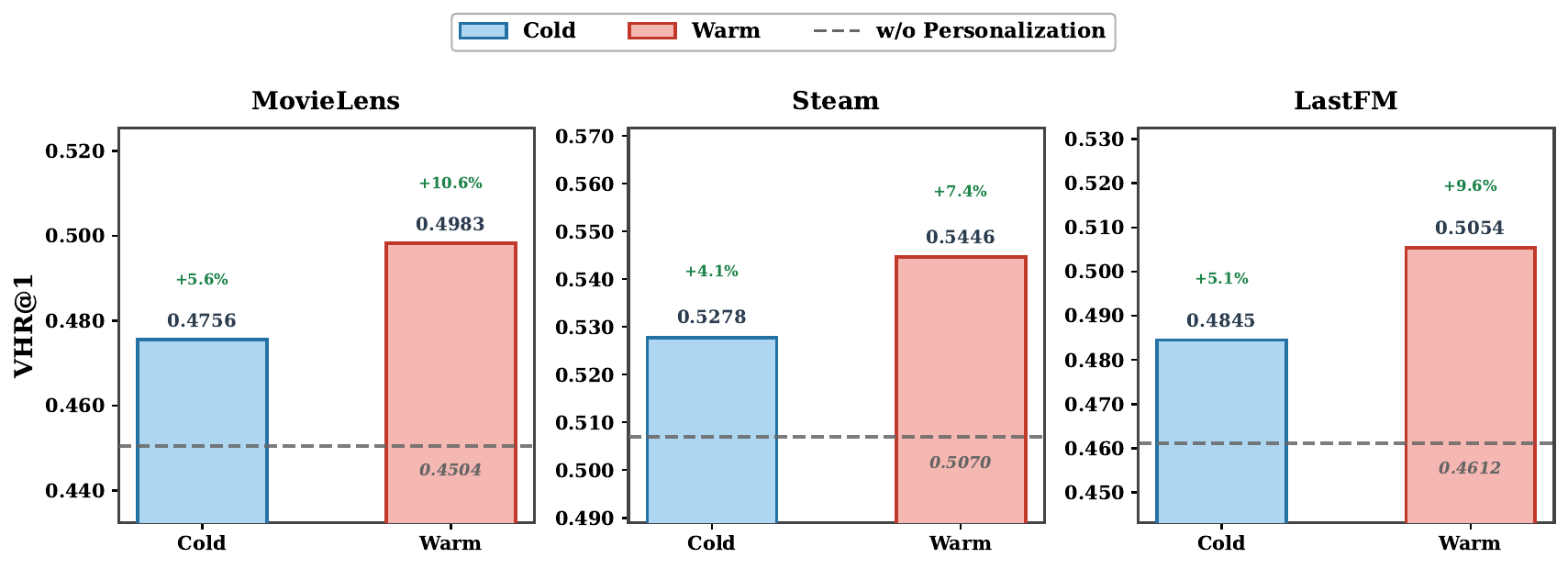}
\caption{Cold-start analysis. Bars represent \ourmodel's VHR@1 under two user groups; the dashed line indicates w/o Personalization from Table~\ref{ablation}. Percentages denote relative improvement over w/o Personalization.}
\label{fig:coldstart}
\end{figure}

Since this analysis aims to isolate the effect of our personalized prompt injection module, we compare \ourmodel\ with its non-personalized ablation variant (w/o Personalization) under different data sparsity conditions, rather than re-evaluating all baselines under each history-length bucket.
Results are shown in Figure~\ref{fig:coldstart}.

Based on Figure~\ref{fig:coldstart}, we identify two notable findings:
\textbf{(1) Personalization benefits even under extreme sparsity.}
Even with 3 or fewer interactions (Cold group), the LLM-based preference extraction already captures meaningful behavioral signals (such as genre preferences or interaction style), yielding gains over the non-personalized variant across all datasets.
This validates the effectiveness of leveraging the LLM's deep semantic understanding for user preference modeling under data sparsity: unlike traditional collaborative filtering methods that require dense interaction matrices, our approach can extract high-level preference patterns from minimal textual evidence.
\textbf{(2) Personalization benefit scales with interaction richness.} The relative improvement over w/o Personalization grows from Cold to Warm, and the Warm group performance slightly exceeds the overall results reported in Table~\ref{tab:compare}, which is expected since the full-dataset results represent a weighted average of both Cold and Warm users.
This confirms that the preference extraction mechanism degrades gracefully as data becomes sparser, making \ourmodel\ suitable for real-world deployment where users span a wide spectrum of activity levels.

\subsection{Semantic Drift Visualization}
\label{sec:drift_vis}

To provide direct empirical evidence for the semantic drift phenomenon and validate the effectiveness of our drift constraint, we visualize the training dynamics of the learnable prompts with and without the constraint.
Specifically, we track two quantities throughout training: (1) the average $\ell_2$ distance between each prompt vector and its initialization anchor (\ie the drift distance), and (2) the ValidRatio on the validation set evaluated at each checkpoint.

\begin{figure*}[t]
\centering
\includegraphics[width=0.95\linewidth]{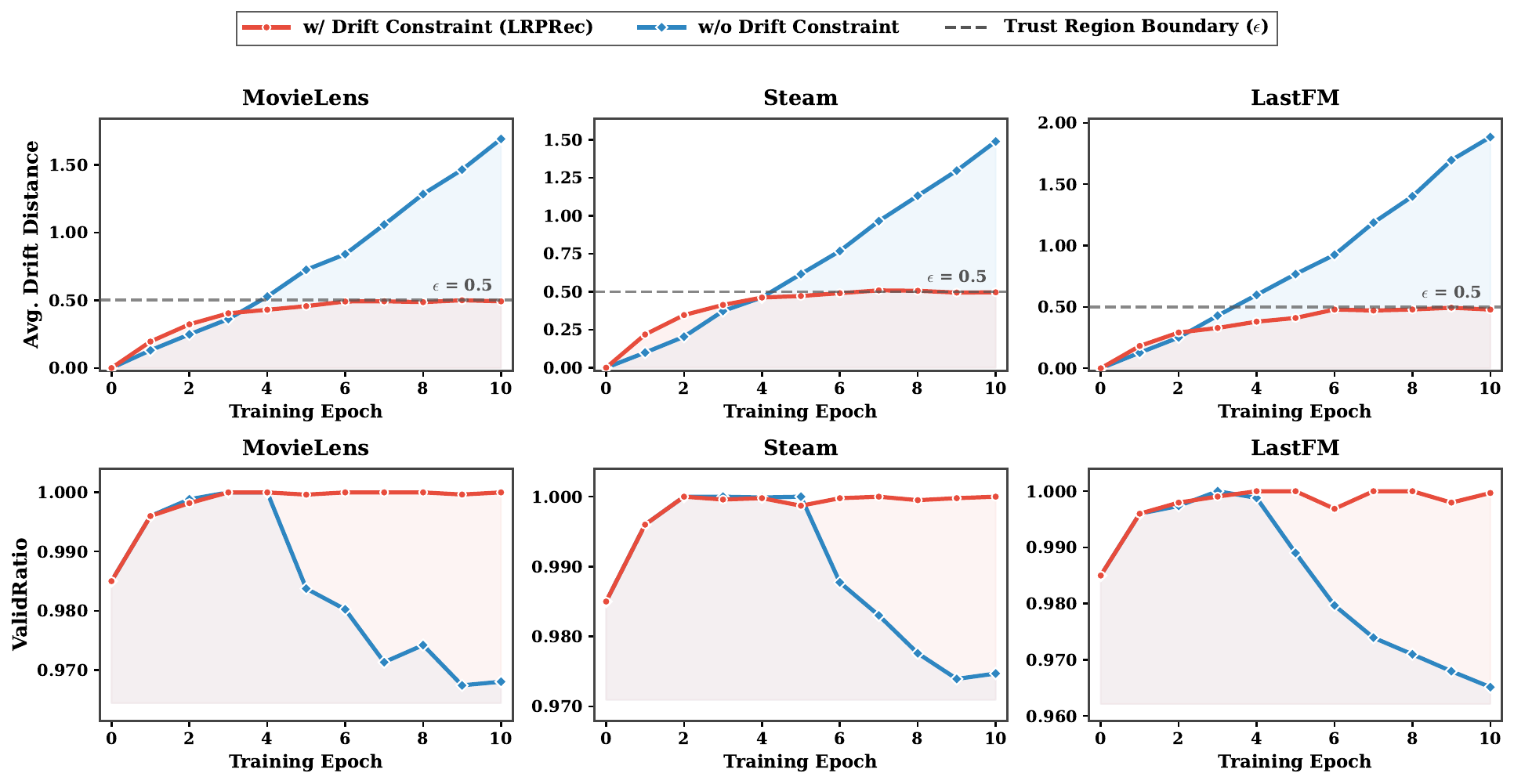}
\caption{Semantic drift visualization during training. Top row: average $\ell_2$ distance between learned prompts and their initialization anchors (drift distance) over training epochs. Bottom row: ValidRatio evaluated at each epoch. The dashed line in the top row indicates the trust region boundary $\epsilon = 0.5$. Without the drift constraint (blue), prompts drift unboundedly and ValidRatio degrades; with the constraint (red), prompts are effectively bounded and ValidRatio remains at 1.0.}
\label{fig:drift_vis}
\end{figure*}

Figure~\ref{fig:drift_vis} presents the results across all three datasets.
The top row shows the drift distance trajectories, and the bottom row shows the corresponding ValidRatio curves.

\textbf{Drift Distance Analysis (Top Row).}
Without the drift constraint (blue curves), the prompt vectors drift continuously away from their initialization anchors throughout training.
The drift distance grows monotonically and exceeds the trust region boundary $\epsilon = 0.5$ around epoch 4--5, eventually reaching large values by the end of training.
This confirms that unconstrained gradient-based optimization indeed pushes the prompts far from the manifold of meaningful token embeddings.
In contrast, with the drift constraint (red curves), the drift distance rises quickly during the initial optimization phase but is effectively bounded near $\epsilon$, demonstrating that the hinge-style regularization successfully confines the prompts within the trust region while still allowing meaningful optimization.

\textbf{ValidRatio Analysis (Bottom Row).}
The ValidRatio curves reveal a clear causal relationship between drift distance and generation quality.
With the drift constraint, ValidRatio stabilizes at 1.0 after a brief warm-up phase, indicating that the LLM consistently produces valid outputs when the prompts remain within the effective semantic space.
Without the constraint, ValidRatio begins to degrade once the drift distance exceeds the trust region boundary. Notably, the onset of degradation (around epoch 4--5) coincides precisely with the point where the unconstrained drift distance surpasses $\epsilon$.
By the end of training, ValidRatio drops noticeably on all three datasets, consistent with the ablation results in Table~\ref{ablation} (w/o Drift Constraint).

\textbf{Key Insight.}
This visualization provides intuitive evidence for the mechanism underlying our drift constraint: the trust region boundary $\epsilon$ acts as a critical threshold. Within this boundary, the LLM can interpret the prompt context meaningfully; beyond it, the prompts enter regions that the LLM was never trained to process, leading to degraded generation quality.
The constraint thus serves as a spatial safeguard that balances expressiveness (allowing prompts to improve upon the discrete template) with stability (preventing departure from the LLM's effective semantic space).

\section{Conclusion}

In this paper, we proposed \ourmodel, a learnable prompting framework for LLM-driven sequential recommendation that addresses the brittleness caused by discrete natural-language prompt formulations.
Our work is motivated by the observation that existing LLM-driven recommenders are highly sensitive to prompt wording, and that naively applying standard prompt tuning to recommendation faces two fundamental challenges: lack of personalization and semantic drift risk.

To overcome these challenges, \ourmodel decomposes the prompt into instance-invariant instruction components and instance-dependent evidence, replaces the brittle discrete instruction templates with continuous learnable vectors initialized from handcrafted templates, and introduces two complementary mechanisms.
First, we propose a personalized prompt injection mechanism that encodes user behavioral signals into a preference embedding via the frozen LLM and directly adds it to the shared instruction prompts, enabling user-level adaptation while maintaining parameter efficiency.
Second, we introduce a semantic drift constraint that regularizes the shared learned prompts to remain within a trust region around their initialization anchors, preventing gradient-based optimization from pushing prompts into regions that the LLM cannot interpret meaningfully.
By applying the drift constraint to the shared component while allowing personalization to operate additively, these two mechanisms are decoupled so that stability and expressiveness can be optimized without mutual interference.

Extensive experiments on three datasets demonstrate that \ourmodel consistently outperforms strong baselines from multiple categories, including ID-based sequential recommenders, multimodal methods, LLM-driven methods, and prompt learning enhanced methods, while maintaining perfect output validity across all datasets.
Ablation studies confirm the necessity of each component, and generalization analysis shows that \ourmodel remains robust regardless of which handcrafted template is used for initialization, effectively eliminating the need for manual tuning of background and task inference templates.
The semantic drift visualization further provides direct empirical evidence for the drift phenomenon and validates the effectiveness of our trust-region constraint.

Future work could explore extending learnable prompting to more realistic settings with dynamic candidate generation, larger candidate pools, cross-domain transfer, and online feedback to further improve adaptivity and scalability.

\clearpage

\begin{acks}
This work was supported by the National Natural Science Foundation of China (No. 624B2131).
\end{acks}

\bibliographystyle{ACM-Reference-Format-unsrt}
\balance
\bibliography{sample-base}

\end{document}